\documentclass[useAMS,usenatbib]{mn2e}

\input psfig.sty
\def\etal{et al.}

\title[The shapes of galaxies in the Sloan Digital Sky Survey]{The shapes of galaxies in the Sloan Digital Sky Survey}
\author[Nelson D. Padilla and Michael A. Strauss]{Nelson D. Padilla$^{1}$\thanks{E-mail:
npadilla@astro.puc.cl} and Michael A.
Strauss$^{2}$\\
$^{1}$Departamento de Astronom\'\i a y Astrof\'\i sica, Pontificia Universidad Cat\'olica de Chile, Santiago, Chile\\
$^{2}$Department of Astrophysical Sciences, Princeton University,
Princeton, NJ 08544 USA}
\begin{document}

\date{Accepted 2008 May 17. Received 2008 April 28; in original form 2008 February 6}

\pagerange{\pageref{firstpage}--\pageref{lastpage}} \pubyear{2008}

\maketitle

\label{firstpage}

\begin{abstract}
We determine the underlying shapes of spiral and elliptical galaxies
in the Sloan Digital Sky Survey Data Release 6 from the observed
distribution of projected galaxy shapes, taking into account the
effects of dust extinction and reddening.  We assume that the
underlying shapes of spirals and ellipticals are well approximated by
triaxial ellipsoids.  The elliptical galaxy data are consistent with
oblate spheroids, with a correlation between luminosity and
ellipticity: the mean values of minor to middle axis ratios are
$0.41\pm0.03$ for $M_r \approx -18$ ellipticals and $0.76\pm0.04$ for 
$M_r \approx -22.5$ ellipticals.  Ellipticals show almost no dependence of axial ratio on
galaxy colour, implying a negligible dust optical depth.

There is a strong variation of spiral galaxy shapes with colour
indicating the presence of dust.
The intrinsic shapes of spiral
galaxies in the SDSS-DR6 are consistent with flat disks with a mean
and dispersion of thickness to diameter ratio of $(21\pm 2)\%$,
and a face-on ellipticity, $e$, of $\ln(e)=-2.33\pm0.79$.
Not including the effects of dust in the model leads to
disks that are systematically rounder by up to $60$\%.
More luminous spiral galaxies tend to have thicker and rounder disks
than lower-luminosity spirals.
Both elliptical and spiral galaxies tend to be rounder for larger
galaxies.

The marginalised value of the edge-on $r$-band dust extinction $E_0$
in spiral galaxies
is $E_0 \simeq 0.45$ magnitudes for galaxies of median colours,
increasing to $E_0=1$ magnitudes for $g-r>0.9$ and $E_0=1.9$
for the luminous and most compact galaxies, with half-light radii $<
2\,h^{-1}$kpc.
\end{abstract}

\begin{keywords}
galaxies: structure, galaxies: general, galaxies: fundamental parameters, surveys
\end{keywords}

\section{Introduction}

The quantitative study of intrinsic galaxy shapes started with Hubble
(1930), who measured the projected axial ratios of elliptical galaxies
when classifying them into what would later become the Hubble
sequence.  Using the projected axial ratios measured from photographic
plates of 254 spiral galaxies from the Reference Catalogue of Bright
Galaxies (de Vaucouleurs \& de Vaucouleurs 1964), Sandage, Freeman \&
Stokes (1970) concluded that the disks of spiral galaxies were
circular, with a disk thickness (defined as the ratio of disk height
to diameter) of $\gamma=0.25$.  Later estimates from photographic
plate surveys were performed by Binggeli (1980), Benacchio \& Galletta
(1980), and Binney \& de Vaucouleurs (1981), who concluded that
galactic disks were consistent with almost circular ellipses, with a
mean ellipticity of $\epsilon=0.1$.  These results, based on small
samples of galaxies, have been superseded in recent years by much
larger studies from CCD imaging and scans of wide-field photographic
surveys (Fasano \& Vio 1991).  Lambas, Maddox \& Loveday (1992)
analyzed a sample of $\sim 13,000$ APM galaxies, and 
found that the distribution of ellipticities was well fitted by a
one-sided Gaussian distribution centred on $\epsilon=0$ with
a dispersion of $\sigma_{\epsilon}=0.13$ and a 
mean of $\langle\epsilon\rangle=0.1$.  Rix \& Zaritsky (1995) studied 
a sample of kinematically selected face-on spiral galaxies in more
detail, finding a typical ellipticity  of $\epsilon=0.045$ in the
galactic disk potential. 

Spatially resolved observations of internal kinematics can sort out the
three-dimensional shape of a galaxy
(Binney 1985; Franx \etal\ 1991; Statler 1994ab, Statler \& Fry 1994; Bak
\& Statler 2000; Statler, Lambright, \& Bak 2001).  Andersen \etal\
(2001) and Andersen \& Bershady (2003) applied this method to 24 
largely face-on spirals and found a mean ellipticity of $\langle
\epsilon \rangle =0.076$, similar to that of Rix \& Zaritsky. However, in both cases
the selection of face-on objects may have introduced systematic biases
in the sample.  Future work with the SAURON spectrograph (Bacon
\etal\ 2001; de Zeeuw \etal\ 2002) will allow detailed
three-dimensional models to be created for a much larger number of galaxies.

Taking advantage of the large number of galaxies with high-quality
photometry and shape measurements in the Sloan Digital Sky Survey
(SDSS; York \etal\ 2000), Ryden (2004) selected a sample of spiral
galaxies from the SDSS Data Release 1 (DR1, Abazajian \etal\ 2003),
chosen to minimise systematics due to seeing.  She found that the
distribution of galactic disk ellipticities can be well fit by a
Gaussian distribution in $\ln \epsilon$ with a mean of $-1.85$ and a
standard deviation of 0.89. Vincent \& Ryden (2005) extended this work
using the 
SDSS Data Release 3 (Abazajian \etal\ 2005), and fit the distribution
of axis ratios of both ellipticals and spirals to triaxial models.
Assuming a uniform triaxiality (i.e. all galaxies are either prolate,
triaxial or oblate),
they found that both spiral and elliptical distributions are
consistent with oblate spheroids.  Moreover, high luminosity elliptical
galaxies show rounder shapes than do lower-luminosity ellipticals.  

Elliptical galaxies were once believed to be
axisymmetric oblate spheroids, until it was discovered that
their rotation velocities were insufficient to support such a geometry
(Bertola \& Capaccioli 1975).  Binney (1976) suggested
that ellipticals could be well described by a triaxial ellipsoid
but Davies \etal\ (1983) found that small ellipticals
are better fit by oblate spheroids.
This variety of intrinsic shapes for elliptical galaxies makes it difficult to
obtain their intrinsic shapes using only their apparent images; when this approach
is used on large numbers of elliptical galaxies, it is often necessary to assume
a triaxiality as in Vincent \& Ryden (2005), or to use the misalignment
between the internal isophotes of individual elliptical galaxies as suggested
by Binney \& Merrifield (1987).
The study of intrinsic shapes of spheroids has been recently extended
to bulges in spiral galaxies by M\'endez-Abreu et al. (2008); bulge shapes are found to be
consistent with a mean axial ratio in the equatorial plane of $\left<B/A\right>=0.85$.

Astronomers as early as Holmberg (1958) realised that the
shape distribution of spiral galaxies is affected by the presence of
dust.  Optically thick dust obscuration
aligned in the rotational plane of spirals will cause
edge-on objects to appear systematically fainter, and thus they will be
under-represented in magnitude-limited samples, biasing the estimates
of intrinsic galaxy shapes.  The dust extinction of galaxies is
important for understanding the true luminosities of galaxies, the
distribution of ISM in galaxies, and the relationship between optical
and infrared emission from galaxies (for reviews, see Davies \&
Burstein 1995 and Calzetti 2001), and studies of the brightness of galaxies as a
function of axial ratio should allow the effects of dust to be
quantified.  Valentijn (1990) studied the shapes and surface brightnesses of $16,000$ 
galaxies from digitised photographic plates, and interpreted the data
as indicating an optically thick component in disk galaxies, extending
well beyond the apparent optical extent of the galaxy.  
Burstein, Haynes \& Faber (1991) and Choloniewski (1991) however, showed that Valentijn's
results were due in part to selection effects, and found 
that the diameters of galaxies were independent of
inclination; see Davies \etal\ (1993) and Valentijn (1994) for further
discussion of these issues. 
Peletier \& Willmer (1992) expanded on the effects of selection biases
with inclination, and emphasised that that the dust opacity may depend
on galaxy luminosity.  Tully \etal\ (1998), for example, found a 1.3
mag difference in the R band between face-on and edge-on luminous
galaxies, but found a negligible effect for intrinsically faint
galaxies.  
Holwerda \etal\ (2005a, 2005b) used a more direct method for obtaining
the opacities of spiral disks, consisting of measuring the
number of field galaxies seen through galactic disks using images from
the Hubble Space Telescope WFPC2 archival data.  This method 
had previously been applied to ground-based data by many other authors, including
Zaritsky (1994), Nelson Zaritsky \& Cutri (1998), and Keel \& White (2001).
Valotto \& Giovanelli (2005) followed a different approach to derive the
dust extinction in galaxies, using the inner part of the rotation curves of spiral galaxies.

More recently, a number of groups have studied the variation of galaxy properties
with the inclination angle with respect to the line-of-sight, or more directly with 
projected galaxy shapes, to draw conclusions regarding dust extinction in spiral galaxies.
Shao \etal\ (2007) measured dust extinction in spiral SDSS-DR2 
(Abazajian \etal\ 2004) galaxies by studying
the luminosity function of galaxies with different inclination angles, and using the intrinsic
galaxy shapes as inferred from the distribution of projected axis ratios.  They interpret the decrease 
in characteristic LF luminosity ($L^*$) with increasing inclination as an effect of dust extinction,
where the disk optical depth is roughly proportional to the cosine of the inclination angle.
However, they did not take into account the influence of dust on the projected shapes of galaxies.
Unterborn \& Ryden (2008) also study the variation of the luminosity function with
inclination using a subsample of $\sim 78,000$ galaxies from 
the SDSS Data Release 6 (DR6, Adelman-McCarthy \etal\ 2008), finding
similar results for the dependence of extinction on projected shape.
They use this to 
define an extinction-unbiased sample of spiral galaxies for which they estimate
intrinsic shapes.  Even though their results indicate that these galaxies are consistent
with flattened disks as was found by previous authors (e.g. Ryden 2004), 
the definition of the sample makes it difficult to compare their
results with previous estimates.   Maller \etal\ (2008) study
the variations 
of galaxy properties with inclination and derive extinction
corrections using the NYU-VAGC
(Blanton \etal\ 2005), which combines data from SDSS and the Two-Micron All Sky Survey 
(Skrutskie \etal\ 2006).  
The median extinction over their whole sample (all morphological
types) is $0.3$ magnitudes in 
the $g$-band.  Finally, Driver et al. (2007) also study the dependence of the
luminosity function with inclination, by decoposing their sample of galaxies in
the Millenium Galaxy Catalogue (Liske et al., 2003, Driver et al., 2005)
into bulge and disk components, and are able to deduce the residual face-on attenuation.

Dust has also been found in elliptical galaxies. Ebneter, Davis \& Djorgovski
(1988) used colour maps to find evidence of dust in more than $30\%$ of
their sample of elliptical galaxies; $\simeq 2.5\%$ of the galaxies
showed evidence for a dusty disk. 
However, the amount of dust in ellipticals is rather smaller than in spirals.  For instance,
Knapp \etal\ (1989) found that an elliptical galaxy contains between $1$ and $10\%$ of
the dust content present in a spiral galaxy of similar luminosity (see also Leeuw \etal\
2004, Krause \etal\ 2003, Goudfrooij 2000).  Far-infrared observations of elliptical
galaxies by Temi \etal\ (2004) also place constraints on the mass of dust in ellipticals in
the range $M_{dust}=10^5-10^7\,h^{-1}$M$_{\odot}$,
where $h$ is the Hubble constant in units of $100\, \rm
kms^{-1}\,Mpc^{-1}$.  This mass is $\sim 10^{-6}$ of the stellar
  mass, a much smaller fraction than seen in spiral galaxies, where
  the fraction is of order $5\times 10^{-3}$
(see for instance Stevens, Amure \& Gear, 2005).

This paper will use the SDSS DR6 to derive the
intrinsic three-dimensional shapes of spiral and elliptical galaxies
using the apparent photometric shapes of galaxies. 
We will include an
in-depth analysis of the effects of dust on the distribution of
apparent shapes of spiral galaxies, which will explain most of the
trend seen in the distribution of spiral shapes with luminosity and
colour.  Our analysis weights galaxies by the inverse of the volume out to which they
can be seen, thus simulating a volume-limited catalogue, and allowing us
to use the full sample of galaxies available in the SDSS DR6.  Furthermore, the
large number of galaxies present in this sample allows us to study the dependence
of the intrinsic shapes of galaxies with luminosity, colour and physical size.
Throughout this paper we assume a standard $\Lambda$CDM cosmology, with
matter density parameter $\Omega_m=0.3$ and a cosmological constant
corresponding to $\Omega_{\Lambda}=0.7$.

This paper is organised as follows.  In Section \ref{sec:gx} we will briefly
describe the SDSS DR6 galaxies, and the parameters we consider when measuring
the distribution of shapes.  Section \ref{sec:shapes} explains our
methodology, including our model for the effects of dust on the
observed distribution of axis ratios.  Section \ref{sec:results} shows
our results, and
Section \ref{sec:discussion} summarises the main conclusions drawn from this work.

\section{The SDSS Galaxy Sample}
\label{sec:gx}

We select the $\sim 585,000$ galaxies from the $r < 17.77$ magnitude-limited main
spectroscopic galaxy sample of the SDSS (Strauss \etal\ 2002) from the
SDSS DR6 (Adelman-McCarthy \etal\ 2008).  
The SDSS imaging data consist of CCD imaging data in five photometric bands ($ugriz$, 
Fukugita \etal\ 1996), taken with a drift-scan camera (Gunn \etal\
1998) on a dedicated wide-field 2.5m telescope (Gunn \etal\ 2006).
The properties of all detected objects in the images are measured
(Lupton \etal\ 1999; Stoughton \etal\ 2002), and are calibrated
astrometrically (Pier \etal\ 2003) and photometrically (Smith \etal\
2002; Ivezi\'c \etal\ 2004; Tucker \etal\ 2006).  
We K-correct the galaxy magnitudes using V3.2 of the code described in
Blanton \& Roweis (2006).

The image of each galaxy in the SDSS sample is fit to two-dimensional
models of a de Vaucouleurs (1948) surface profile and an exponential
profile, each convolved with the PSF of the image.
This fitting procedure provides a measurement of the model axial
ratios ($b/a$) of each galaxy image in a way robust to seeing, as well
as the effective radius and position angle of the galaxy.  The SDSS
image pipeline also fits an ellipse to the 25 mag/arcsec$^2$ isophote
of each galaxy, and determines the so-called adaptive moments
(Bernstein \& Jarvis 2002);
while the axis ratios via these statistics are generally in good
agreement with those from the model fits, they are affected by seeing,
and thus tend to yield systematically rounder shapes than do the model
fits.

\begin{figure*}
\begin{picture}(430,340)
\put(0,-35){\psfig{file=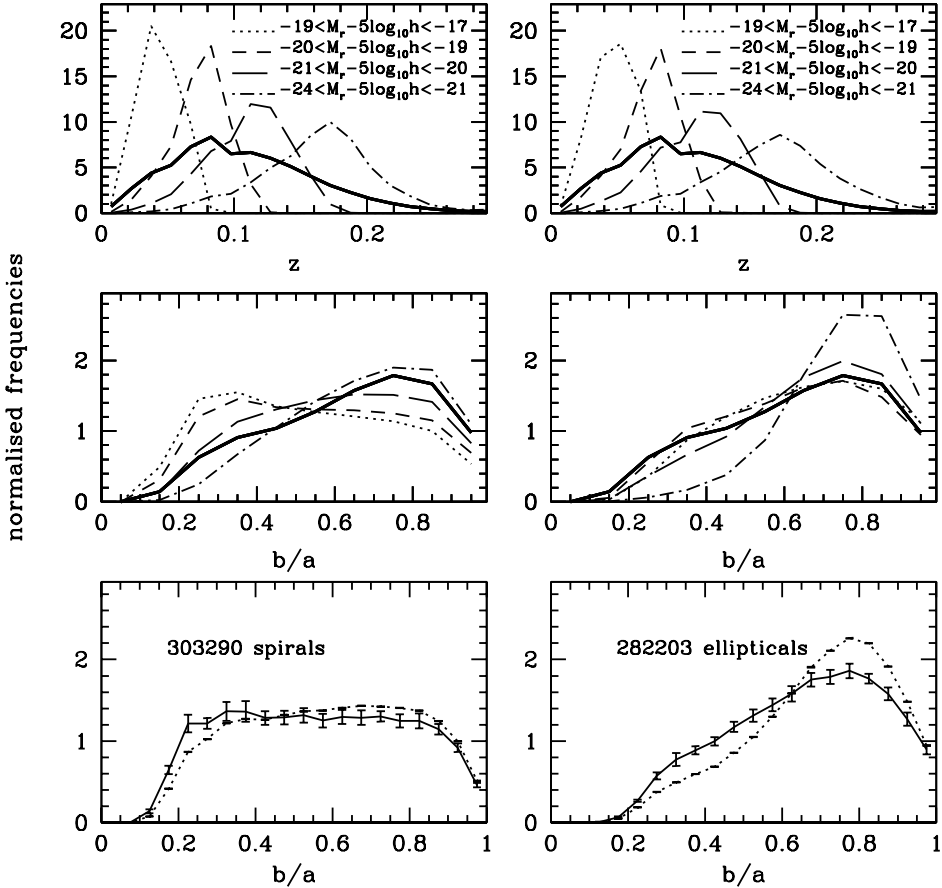,width=14.cm}}
\end{picture}
\caption{
Left panels show distributions obtained for spiral galaxies, right panels for elliptical galaxies.
Top panels: distribution of galaxy redshifts, normalised so that the area under each curve is 
unity.  The thick solid line corresponds to all the galaxies in the SDSS-DR6.  
The thin lines show the 
distribution of axis ratios for galaxies in different bins of absolute magnitude as
indicated in the key, and for spiral and elliptical galaxies separately. Middle
panels: normalised distribution of axis ratios for the same samples of galaxies as in the top
panels; the thick line corresponds to the full sample of galaxies.  
Bottom panels: Distribution of axis ratios, summed over all luminosities. The solid lines show the results when a 
$1/V_{max}$ weight is applied to each galaxy.  The dotted lines show the results with 
no weighting.  Errors are calculated using the Jack-knife technique.
}
\label{fig:badr3figs}
\end{figure*}

\begin{figure*}
\begin{picture}(430,370)
\put(0,-20){\psfig{file=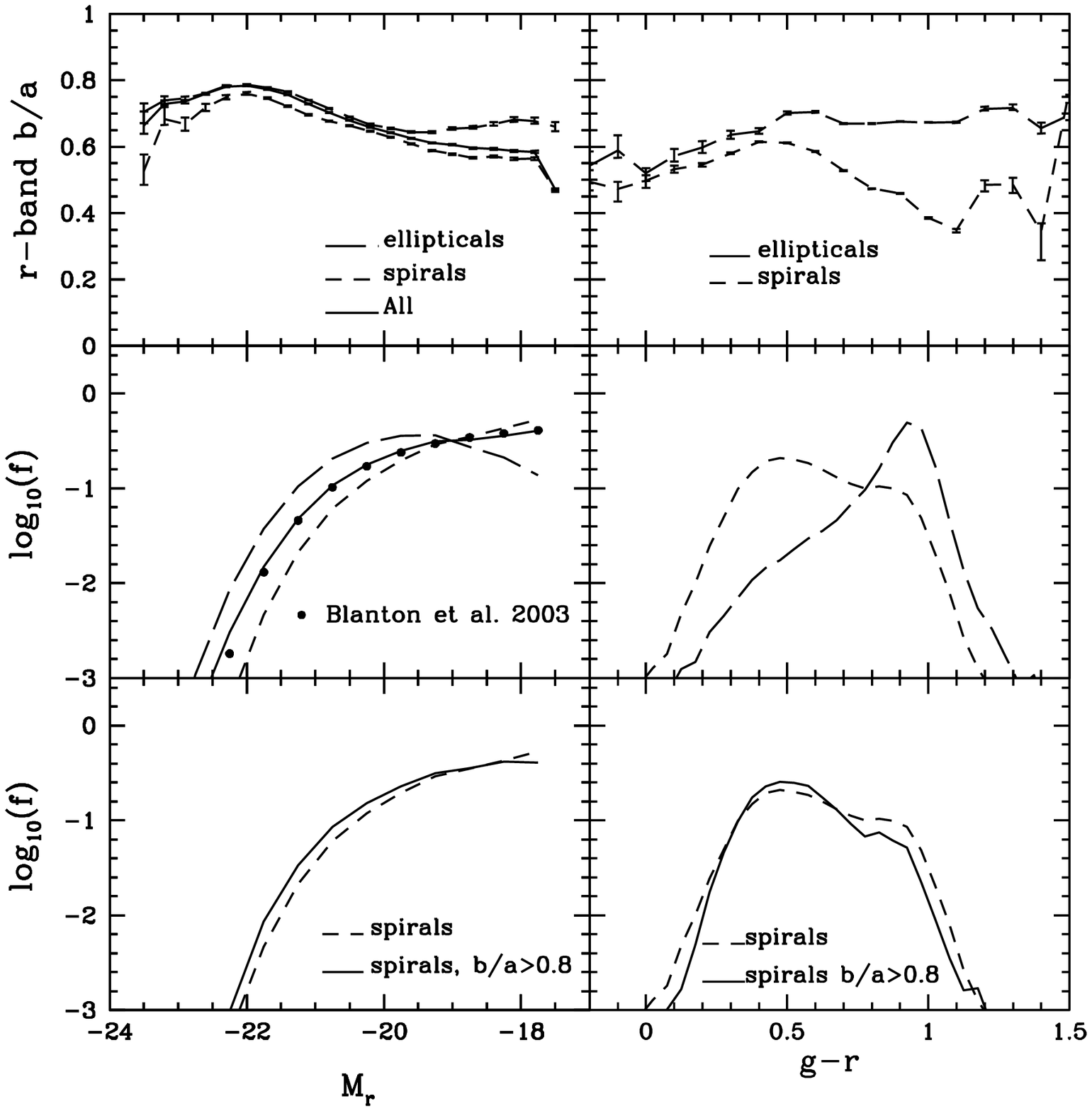,width=14.cm}}
\end{picture}
\caption{
  Top-left: median axis ratios of spiral (short dashed) 
and elliptical galaxies (long dashed) 
as a function of $r$-band absolute magnitude; errors are calculated
using the Jack-knife technique.  The middle-left
panel shows the distributions of absolute magnitudes (luminosity functions) as short and long dashed
lines.  The filled circles show the best Schechter fit to SDSS galaxies from Blanton
\etal\ (2003a) normalised so that the area under the curve is equal to one.
The bottom-left panel shows the luminosity functions for all spiral galaxies, and for
the subset of face-on spiral galaxies ($b/a>0.8$) which we use as our
unextincted luminosity function. Right: Median axis ratio and absolute
magnitude as a function of $g-r$ colour for spirals and
ellipticals separately.  We do not show results for the full sample to
improve clarity.  
}
\label{fig:absmdr3}
\end{figure*}

We will infer the three-dimensional shapes of spiral and elliptical
galaxies separately.
Park \& Choi (2005) presented a very accurate way to determine SDSS galaxy
morphologies using colour gradients that could be used to separate the DR6 catalogue
into spiral and elliptical galaxies, but this would require
analyzing the images of each individual galaxy separately.  
We use an alternative method:
in fitting the exponential and de Vaucouleurs models, the SDSS imaging
pipeline also asks for the best linear combination of these models
(Abazajian \etal\ 2004), as quantified by the parameter $\rm
fracDeV$.  We use this parameter to distinguish spiral galaxies
(${\rm fracDeV}<0.8$) from ellipticals (${\rm fracDeV}\ge0.8$).
  It is not possible using these techniques to separate out
  lenticular or $S0$ galaxies; the definition of this morphological
  type is difficult using photometric data, out to the redshifts explored in this work.
The axis ratios from the
exponential and de Vaucouleurs models are in excellent agreement, 
independent of the value of fracDeV, 
with a scatter of about 0.05 around the identity line, but we adopt the
axis ratios from the exponential fit when 
${\rm fracDeV}<0.8$, 
and the de Vaucouleurs  parameters otherwise.
All our analyses are carried out using 
model DeVaucouleurs or exponential $r$-band magnitudes and $g-r$ colours, depending on the
galaxy type.

\section{The intrinsic shapes of galaxies}
\label{sec:shapes}

In this section we will measure the distribution of projected axis
ratios and present the model that will allow us to infer
their intrinsic shapes.

\subsection{Distributions of projected axis ratios of elliptical and spiral galaxies}

Figure \ref{fig:badr3figs} shows distributions of spiral galaxies on the left panels,
and of elliptical galaxies on the right panels.  The top panels show
distributions of galaxy redshifts, and the middle and lower panels the distributions of projected
axis ratios.  In addition, the thick solid lines in the top and middle panels show the distributions
of redshifts and projected axis ratios for the full sample of galaxies in the SDSS-DR6.
In these panels, the thin lines illustrate
the variation of redshift and projected axis ratio distributions as the typical 
galaxy luminosity of the sample is increased.  More luminous
galaxies tend to show rounder apparent shapes (axial 
ratios closer to unity), suggesting that the intrinsic shapes of
galaxies are a function of luminosity.  This is a flux-limited sample,
and thus has a strong correlation between luminosity and redshift,
as shown in the top panel.  With this in mind, we simulate a
volume-limited measurement of apparent shapes by simply weighting each
galaxy by $1/V_{max}$, where $V_{max}$ is the volume corresponding to the 
maximum distance out to which a galaxy of a given apparent magnitude 
enters the flux-limited catalogue, taking into account K-corrections\footnote{
The advantage of using the $1/V_{max}$ weight is the larger sample size and 
the increased range of luminosities that can be explored; using a true volume-limited sample
would restrict our analysis to a sample almost $\sim 10$ times smaller composed only
by intrinsically bright galaxies.  }. 
As a result of these procedures, our final samples of spirals
and ellipticals contain a total of $282203$ and $303390$ galaxies, respectively.
There is an additional effect for spiral galaxies, whereby the
internal extinction will cause edge-on objects to appear fainter than
equally luminous face-on objects.  We thus first volume-limited
ignoring this effect, determined the inclination dependence of the
extinction as described in \S~\ref{ssec:dust} below, and then redefined our volume-limited
sample taking this extinction into account and repeated our analysis.
We found that the inferred extinction differed by only 0.1 mag between
the two analyses, so we did not iterate further.
The solid lines in the bottom panels of Figure~\ref{fig:badr3figs} show 
the resulting distributions of axial ratios for
spiral and elliptical galaxies.
Not surprisingly, the spiral galaxy distribution is 
skewed toward lower axial ratios ($b/a$ values) than are the ellipticals, which are rounder 
with $b/a$ closer to $1$. 

\begin{figure}
\begin{picture}(230,250)
\put(0,-20){\psfig{file=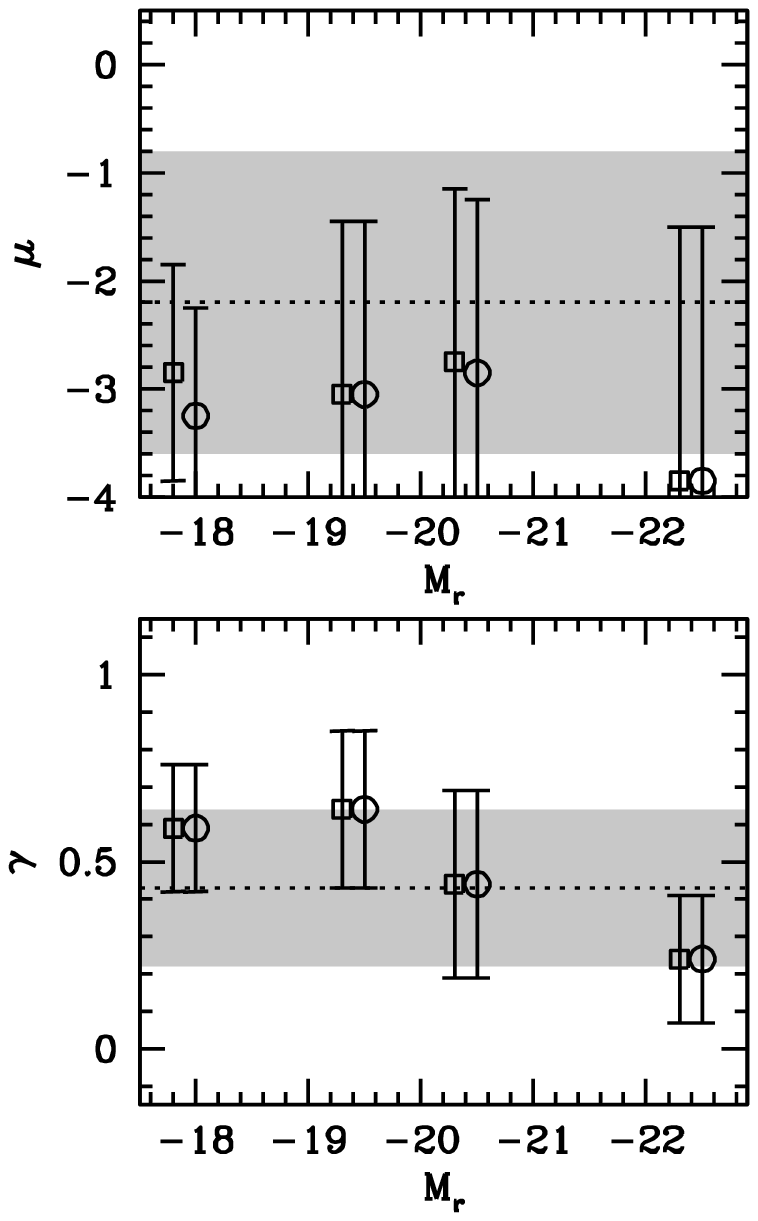,width=14.cm}}
\end{picture}
\caption{
Dependence of the best-fitting parameters on galaxy luminosity, for
elliptical galaxies.  
Squares represent the best fit parameters from the marginalised, one-parameter
probabilities.  
Circles correspond to the best fit parameters in the four-dimensional
parameter space.
Dotted lines and shaded areas 
indicate the best fit parameters for the full sample of elliptical galaxies.
Top panel: variations
in the typical 
$\mu={\rm median} \log(1-B/A)$
axis ratio; the error bars and shaded areas 
do not correspond to uncertainties in the parameters but indicate the best
fitting width $\sigma$ for the Gaussian
distribution of $\mu$ values used in the model.
Bottom panel: same as top panel for 
$\gamma={\rm median} (1-C/B)$.
}
\label{fig:var_devgx}
\end{figure}

\begin{figure}
\begin{picture}(230,240)
\put(0,0){\psfig{file=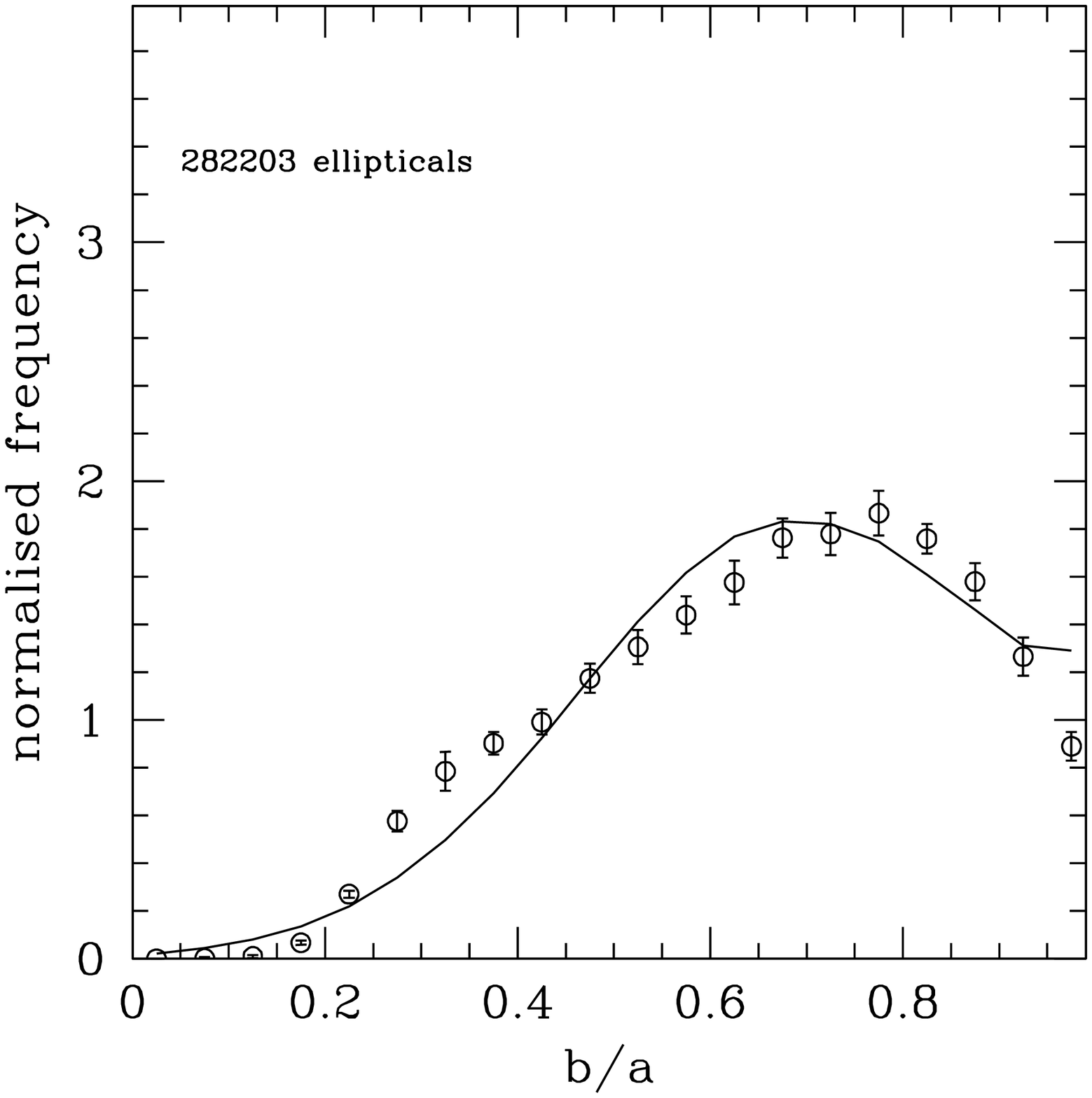,width=8.cm}}
\end{picture}
\caption{
Comparison between the best fit model axis ratio distributions and the
actual measured distributions from elliptical galaxies in the SDSS
DR6.  Errors are calculated using the Jack-knife 
technique.
}
\label{fig:model_dr3_devgx}
\end{figure}

The distribution becomes flatter when using
the $1/V_{max}$ weighting.
In particular, the weighted distribution of $b/a$ values for spiral galaxies is qualitatively similar
to that of Ryden (2004), who found a flat distribution over a wide 
range of $b/a$ values from $b/a=0.2$ to
$0.7$ for a volume limited sample of SDSS-DR2 spiral galaxies.  
We use $1/V_{max}$ weighting in all the analyses that follow.  Galaxies fainter than
$M_r-5\log_{10}(h) = -17$ have very small values of $V_{max}$, and thus tend to
dominate the noise of the estimate of the distribution function; we
therefore drop such low-luminosity objects in what follows.
Throughout this paper, error-bars were calculated using the Jack-knife method.

We model galaxies as
triaxial ellipsoids of major axis $A$, middle axis $B$, and minor axis $C$,
parameterised by two axis ratios, $C/B$ and $B/A$, and we will determine the
distribution of axis ratios of the spiral and elliptical populations separately.  Following 
Ryden (2004), we assume that the distribution of $1-C/B$
of the three-dimensional structure
can be approximated by a Gaussian with mean $\gamma$ (this parameter is
related to $\mu_{\gamma}$ in Ryden 2004, via $\gamma=1-\mu_{\gamma}$) and 
standard deviation $\sigma_{\gamma}$. 
We also assume that there is a log-normal distribution in the quantity 
$\epsilon=\log(1-B/A)$, 
with mean $\mu$ and dispersion $\sigma$.  
Larger values of $\gamma$ and $\mu$ correspond to more elliptical objects in the
$B-C$ and $A-B$ planes, respectively.
Given values of axis ratios drawn from these distributions, 
and a random
viewing angle $(\theta,\phi)$, we compute the resulting apparent axis
ratio, $b/a$ (Binney 1985), using equations $12-15$ from Ryden (2004).
Repeating this multiple times gives a model distribution ($N_{\rm model}(b/a)$) 
which can be compared directly to our measured volume-weighted distributions 
($N(b/a)$).  

This model assumes that the dust extinction in galaxies is
independent of the viewing angle.  However, we can test for the effects of dust by
exploring the dependence of the
shape distribution on absolute magnitude and colour.
The top-left panel of Figure \ref{fig:absmdr3} shows the median $b/a$ 
of spiral and elliptical galaxies as a function of $r$-band absolute
magnitude.  Both spiral and elliptical galaxies tend to be rounder at
higher luminosities.  
Spiral galaxies show a change of $\Delta(b/a)\simeq0.2$ between absolute
magnitude values of $M_r-5\log_{10}(h)=-18.5$ and $-22.5$.  
The variation in $b/a$ for ellipticals
is smaller, $\Delta(b/a)\simeq0.1$.

The top-right panel shows the median $b/a$ as a function 
of $g-r$ colour.  Red spiral galaxies show systematically larger axis
ratios than do blue spirals.  We interpret the colour and luminosity
dependence of spirals as due to the reddening and dimming effects of
dust, which becomes more prominent for edge-on systems.  
Elliptical galaxies show only a mild variation in shape with absolute
magnitude, and no significant effect in colour.  Thus, not
unexpectedly, we see no evidence for a dust layer aligned with the
principal plane of elliptical galaxies,
and assume that the trend in
the apparent shape of ellipticals with luminosity corresponds to a
real variation of their intrinsic shapes.
We study the dependence of projected and intrinsic properties of spiral
and elliptical galaxies in more detail in Section \ref{sec:results}. 

The middle panels of figure \ref{fig:absmdr3} show the luminosity and
colour distribution functions 
(left and right panels, respectively), both estimated using the $1/V_{max}$ estimator, 
for spiral and elliptical galaxies.  The filled circles
in the middle-left panel give the $r$-band luminosity function estimate
from Blanton \etal\ (2003a), with Schechter parameters $M^*-5\log_{10}(h)=-20.44$ and
$\alpha=-1.05$; our results for the full sample (black solid lines) are in excellent
agreement. 
As the effects of dust are more severe for edge-on objects, we also calculate
the luminosity and colour functions for face-on objects ($b/a>0.8$) only.  These are shown
as solid lines for the spiral galaxies in the bottom panels of
this figure; the dashed lines show the luminosity and 
colour functions for all the spiral galaxies
for comparison.  The break in the face-on luminosity function is
shifted towards 
higher luminosities, 
as expected for the sample of
galaxies least affected by dust. 
This measurement of the ``unextincted" luminosity function is 
in agreement with results from Shao \etal\ (2007), Unterborn \& Ryden (2008) and, taking
into account the range allowed by the analysis, with Maller \etal\ (2008). 
On the other hand, the colour function of spiral galaxies 
is shifted to the red due to the effects of dust.
These estimates of unextincted  
luminosity and unreddened colour functions will be needed when we model the
effects of dust, a subject to which we now turn. 

\subsection{Modelling the effects of dust}
\label{ssec:dust}

In this section, we develop a simple model for the impact of a 
planar distribution of dust on the distribution of apparent axis
ratios of spiral galaxies.  We assume no correlation between the dust
column and the physical diameter of the galaxy, an assumption we will
justify a posteriori.  While Unterborn \& Ryden (2008) use the
inclination dependence of the luminosity function to 
define a sample of galaxies not affected by dust (i.e. not biased towards face-on objects),
our shape
fitting solves for the dust effects self-consistently. 

We follow the
following steps to produce the predicted distribution of projected
axis ratios, given our assumed distribution of axis ratios
and our measured luminosity and colour distributions:
\begin{enumerate}
\item We assume that the amount of extinction and reddening are roughly proportional
to the path length of the light through the galaxy.  Therefore, we expect a minimum 
extinction when a galaxy is seen face-on, and an increasing extinction as the 
line-of-sight approaches the plane of the galactic disk.
Similarly, Shao \etal\ (2007), Unterborn \& Ryden (2008) and
Maller \etal\ (2008) all find that the optical depth increases
monotonically with inclination angle. 
The following parametrisation of the angle dependence of dust extinction is not 
intended as a physical model, but as a heuristic guess for the scaling.  Consider an
oblate triaxial galaxy with axis ratios given by $x=B/A$ and $y=C/B$.  The total 
dust extinction as a function of inclination $\theta$ in our model is
\begin{equation}
E(\theta)= \left \{
\begin{array}{l}
E_0(1+y-\cos\theta),  {\rm if} \cos\theta>y \nonumber \\
E_0,  {\rm if} \cos\theta<y 
\end{array} \right.
\label{eq:ext}
\end{equation} 
where $E_0$ is the edge-on extinction in magnitudes in a given band
and 
$y$ is the galaxy height to diameter ratio extracted from a distribution of mean $\gamma$
and width $\sigma_{\gamma}$. 
The same can be assumed for the dust reddening,
\begin{equation}
R(\theta)=  \left \{
\begin{array}{l}
R_0(1+y-\cos\theta), {\rm if} \cos\theta>y \nonumber\\
R_0, {\rm  if} \cos\theta<y
\end{array} \right.
\end{equation} 
where $R_0$ is the edge-on reddening in magnitudes.  In the optically
thin case, we can tie $R_0$ to the extinction via
$E_0 = 2.77 R_0$, as is appropriate for the $r$-band and $g-r$ colour.  We find 
that our model results are not strongly dependent on this parameter and therefore
can be applied to either the optically thin or thick case.

\item We produce an extincted LF defined by
  $$\phi_E(M,\theta)=\phi(M+E(\theta))$$ 
where $\phi(M)$ is the unextincted luminosity function calculated using only face
on galaxies.

\begin{figure*}
\begin{picture}(430,360)
\put(0,-30){\psfig{file=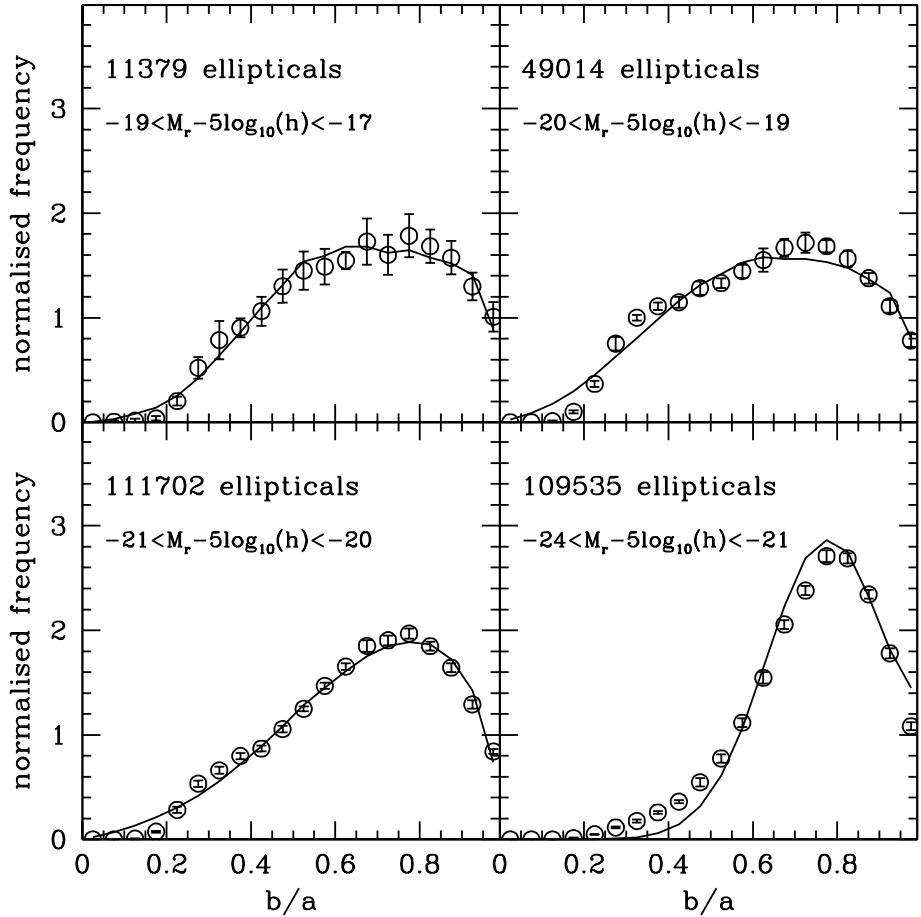,width=14.cm}}
\end{picture}
\caption{
Comparison between the best fit model axis ratio distributions and the
actual measured distributions from elliptical galaxies in the SDSS
DR6 in bins of absolute magnitude.  Errors are calculated using the Jack-knife 
technique.
}
\label{fig:model_dr3_abscut_devgx}
\end{figure*}

We define the ratio, $f_E$, between the number of observed (extincted) and 
intrinsic galaxies of a given luminosity $f_E(M)=\phi_E(M)/\phi(M)$.
Similarly, we define $f_R(g-r)$ as the ratio between the
underlying and reddened distributions of galaxy  colours.
\item We calculate the ratio of the number of galaxies seen at
inclination $\theta$ to the number expected without extinction, by multiplying the effects
of reddening and extinction together,
\begin{equation}
\psi(\theta)=
\frac{\int_{-\infty}^{\infty}\hskip-.08cm\int_{-\infty}^{\infty}\hskip-.08cm f_E(M)f_R({\cal C})\phi_s(M)\phi_s({\cal C})W\hskip-.08cm({\cal C},M){\rm d}{\cal C}{\rm d}M}
{\int_{-\infty}^{\infty}\int_{-\infty}^{\infty}\phi_s(M)\phi_s({\cal C})W({\cal C},M){\rm d}M{\rm d}{\cal C}},
\label{eq:psi}
\end{equation}
where ${\cal C}=g-r$, and the function $W$ contains the correlation between
colour and $M_r$.
We assume that $W$ is Gaussian with mean and dispersion extracted directly from the data; this correlation
is compatible with the results shown in Figures $11$ and $12$ of Blanton \etal\ (2003b).
The sub-index $s$ indicates that the luminosity and colour functions correspond
to a particular subsample of galaxies;  these subsamples are defined using sharp cuts
in the allowed luminosity and colour ranges.  This indicates that $\psi$ depends not only
on the amount of extinction and reddening, but also on the range of luminosities and colours present
in each subsample of galaxies.  Note that Eq. \ref{eq:psi} ignores the
presence of large-scale structure, which is justified given the large solid
angle of the SDSS sample.
\item We then construct the model distribution of apparent axis ratios 
  $N_{\rm model}(b/a)$ as described in Section \ref{sec:shapes}.  
  Rather than selecting the cosine of the viewing angle from a flat
  distribution, we select it from $\psi(\theta)$ as given in
  equation~(\ref{eq:psi}). In the case 
  where $E_0=0$ and $R_0=0$, $\psi(\theta)=1$. 
\end{enumerate}
In general, the effect of dust extinction and reddening is to decrease the 
number of galaxies seen edge-on relative to those that are face-on.
This decrease depends 
strongly on the luminosity and colour functions, as well as on the selected range
of luminosities and colours.  
The above process gives a prediction for the observed projected axis ratio
distribution for a given set of triaxial galaxy parameter distribution
functions and dust properties.  
Note that for a given viewing angle, the model states that the dust
affects the likelihood 
that a galaxy would enter the sample at that viewing angle, but does 
not affect the observed projected axis ratio.  This is a good approximation
as long as the dust is smoothly distributed within individual galaxies.

In the following section, we will
constrain these parameters by fitting these predictions to the
observed axis ratio distribution.

\subsection{Parameter fitting}

We aim to constrain the parameters $\mu$, $\sigma$, $\gamma$ and 
$\sigma_{\gamma}$ which describe the intrinsic shapes of galaxies by
fitting the observed axis ratio distribution.  Spirals and ellipticals
have intrinsically different shapes, and we fit to the two
separately.  For spirals, we also include the effects of dust via the
extinction parameter $E_0$ which in turn defines the reddening $R_0$.  

We define a grid in parameter space $p$ (four parameters for ellipticals, five
for spirals).
Using the parameters of each grid point ${p}_i$, we
generate random three-dimensional axis ratios from the assumed
distribution, observed at a random orientation (modulated by the
effects of dust, as described above).  We then generate
ten independent model distributions of projected axis ratios, each containing as
many galaxies as the sample of galaxies
under analysis.  We take the average of these ten distributions as the
final model distribution, $N_{\rm model}(b/a,\{p\}_i)$, and use the jack-knife errors, 
$\sigma_{\rm jack-knife}(b/a)$, obtained from the
observed distribution to define a $\chi^2$ between the real 
data, $N(b/a)$, and the model,
\begin{equation}
\chi^2(\{p\}_i)=\sum_{b/a\,\rm bins}\left( 
\frac{N_{\rm model}(b/a,\{p\}_i)-N(b/a)}{\sigma_{\rm jack-knife}(b/a)}\right)^2,
\end{equation}
The best fit parameters correspond to the minimum value of $\chi^2$ throughout the 
parameter grid. 
Throughout this analysis, we use a bin size of $\Delta(b/a)=0.1$ in presenting the observed
and model distributions of $b/a$ and when calculating $\chi^2$.

\section{Results}
\label{sec:results}
\subsection{Elliptical galaxies}

The grid of parameters we used for elliptical galaxies is shown in
Table \ref{table:grid1}.  The grids include the full range of
values suggested for these parameters in the literature.
The number of steps for each parameter, shown in the fourth column, 
corresponds to a coarse initial grid; once the best fit parameters are found
we re-do the analysis on a finer grid with twice the 
resolution centred on the best fit parameters.  
We repeat this refinement twice.  Given the lack of colour dependence of the
axis ratio distribution for ellipticals, we have assumed no dust.  
In order to explore the dependence of galaxy shape on luminosity, we
divide the ellipticals by $r$-band absolute magnitude in four bins, with
boundaries given by $M_r-5\log_{10}(h)=
-24,-21,-20,-19$ and $-17$.  

\begin{table}
\centering
\begin{minipage}{80mm}
\caption{Initial parameter grid for Ellipticals}
  \begin{tabular}{@{}cccc@{}}
  \hline
Parameter & Min. Value & Max. Value & Number of steps\\
 \hline
$\mu$ & $-4.05$ & $-0.05$ & $21$ \\
$\sigma$ & $0.4$ & $3$ & $18$ \\
$\gamma$ & $0.09$ & $0.99$ & $19$ \\
$\sigma_{\gamma}$ & $0.01$ & $0.36$ & $12$ \\
\hline
\label{table:grid1}
\end{tabular}
\end{minipage}
\end{table}

\begin{figure}
\begin{picture}(230,220)
\put(0,0){\psfig{file=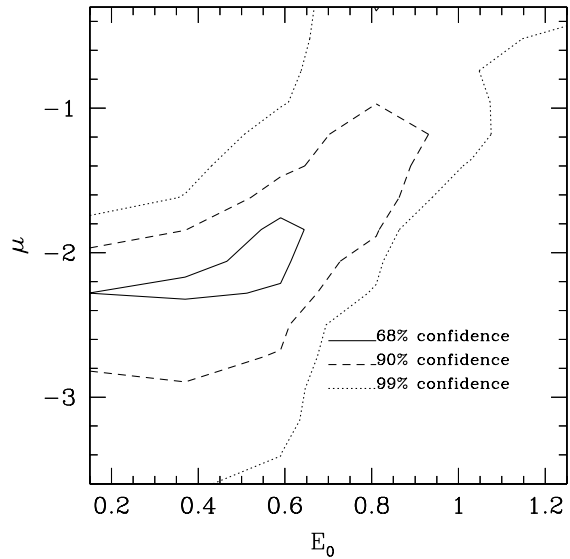,width=8.cm}}
\end{picture}
\caption{
Marginalised likelihood contours in the $\mu$-$E_0$ plane, for the full sample
of spiral galaxies, corresponding to $1-$, $2-$ and $3-\sigma$ confidence levels (shown
as solid, dashed and dotted lines, respectively).  
}
\label{fig:2Dcontours}
\end{figure}

\begin{figure}
\begin{picture}(230,220)
\put(0,0){\psfig{file=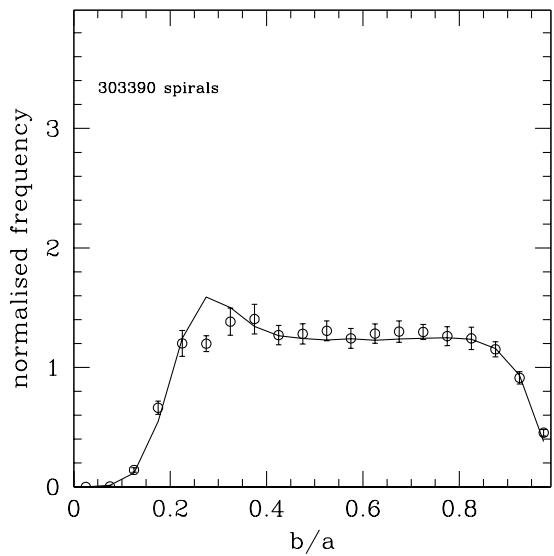,width=8.cm}}
\end{picture}
\caption{
Comparison between the best fit model axis ratio distributions (solid line) and the
actual measured distributions from the full sample of spiral galaxies in the SDSS DR6
(open symbols with error bars).  Errors are calculated using the Jack-knife method.
}
\label{fig:model_dr3_expgx}
\end{figure}

We calculate the marginalised one-parameter likelihoods 
(normalised to the maximum likelihood) resulting from fitting the
observed $N(b/a)$.  
The resulting best-fit parameters are shown in Figure \ref{fig:var_devgx}. 
The ``error bars'' correspond to the best fit widths, $\sigma$ and $\sigma_{\gamma}$,
of the distributions of $\log(1-B/A)$ and $1-C/B$.
As
we guessed from the projected axis ratios, more luminous elliptical
galaxies are consistent with a rounder underlying shape:
the mean axis ratio $\gamma$ changes from $0.6$ to $0.2$ with increasing luminosity (see
Table \ref{table:bestfitsdev} for values of $\gamma$ and estimated errors not shown in the figure).
The quantity $\mu$ is much more constant, with the implied $B/A$
three-dimensional axis ratio varying only slightly, from $\simeq 0.95$ to $\simeq 0.92$.

In this figure, squares show the best-fit parameters corresponding to the marginalised
one-parameter maximum likelihoods.  The open symbols show the best-fit parameters
as obtained from the full parameter space.  In most cases, the two
estimates agree reasonably well.  
Table \ref{table:bestfitsdev} shows the values of the best fit parameters for the 
four subsamples of elliptical galaxies, as well as the maximum likelihood value.  
The reduced $\chi^2$ is $\leq 2$, indicative of a good fit, except for
the highest luminosity subsample and the total sample.  In these case,
our simple model does not give a statistically rigorous good fit, but
it still follows the shape of the observed distribution quite well. 
Figure \ref{fig:model_dr3_devgx} directly compares the observed
axis ratio distribution with the model fits for the full sample of elliptical galaxies.
Figure \ref{fig:model_dr3_abscut_devgx} shows the comparison in each bin of absolute
magnitude; the agreement between the model and observed distributions, while not perfect in every
case, is impressive, and thus the
four-parameter model is adequate to describe the real distributions of axis ratios 
for the galaxy luminosities explored here.

Vincent \& Ryden (2005) used slightly different analysis techniques and a sample a quarter
of the size the one used here, but they found results in very good agreement with our own,
namely that low luminosity elliptical galaxies
are more consistent with prolate spheroids than are high-luminosity ellipticals.

The analysis of the marginalised two-parameter likelihoods indicates that there are little
or no degeneracies between the parameters used to fit the projected shapes of elliptical
galaxies.

\begin{table*}
\centering
\begin{minipage}{120mm}
\caption{Best fit model parameters for elliptical galaxies.}
\begin{tabular}{@{}cccccc@{}}
\hline
Par.  &  Sample 1 &  Sample 2 &  Sample 3 &  Sample 4&All \\
  & \tiny $-17>M_r>-19$ & \tiny$-19>M_r>-20$ & \tiny$-20>M_r>-21$ & \tiny$-21>M_r>-24$ \\
\hline
$\mu$             & $-2.85\pm0.30$ & $-3.05\pm0.20$ & $-2.75\pm0.10$ & $-3.85\pm0.15$&$-2.2\pm0.1$ \\
$\sigma$          &  $1.15\pm0.35$ &  $1.00\pm0.05$ &  $2.60\pm0.15$ &  $2.35\pm0.20$& $1.4\pm0.1$\\
$\gamma$    &  $0.41\pm0.03$ &  $0.36\pm0.06$ &  $0.56\pm0.02$ &  $0.76\pm0.04$ &$0.57\pm0.06$\\
$\sigma_{\gamma}$ &  $0.17\pm0.03$ &  $0.21\pm0.02$ &  $0.25\pm0.02$ &  $0.17\pm0.01$&$0.21\pm0.02$ \\
$\chi^2/$dof        & $0.41$  & $2.0$& $1.72$  & $7.2$ & $6.8$\\
\hline
\label{table:bestfitsdev}
\end{tabular}
\end{minipage}
\end{table*}

\begin{figure*}
\begin{picture}(430,350)
\put(0,-30){\psfig{file=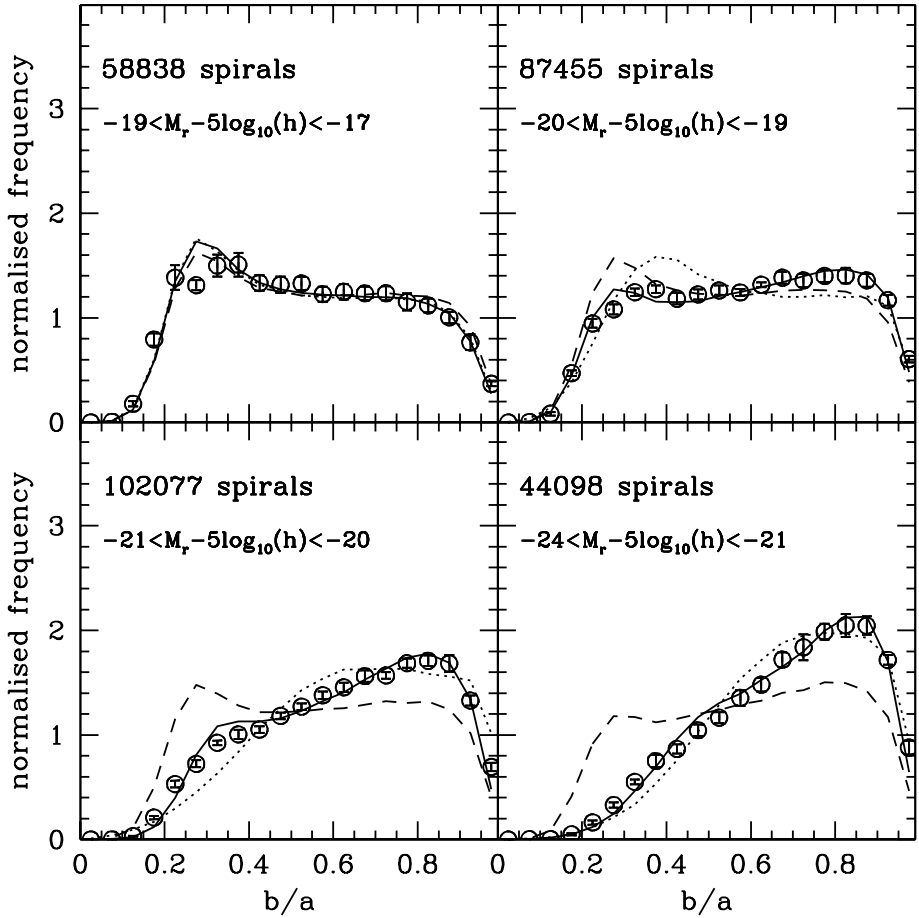,width=14.cm}}
\end{picture}
\caption{
Comparison between the best fit model axis ratio distribution to the full
sample of spiral galaxies (dashed line), and the
actual measured distributions from samples of spiral galaxies with different
luminosities in the SDSS DR6
(open symbols with error bars).  Errors are calculated using the Jack-knife method.
Solid lines show the results of fitting the model to each individual subsample including
dust extinction.  The dotted lines show the best fits when not including dust.
}
\label{fig:model_dr3_abscut_expgx}
\end{figure*}

\begin{figure*}
\begin{picture}(430,415)
\put(0,-30){\psfig{file=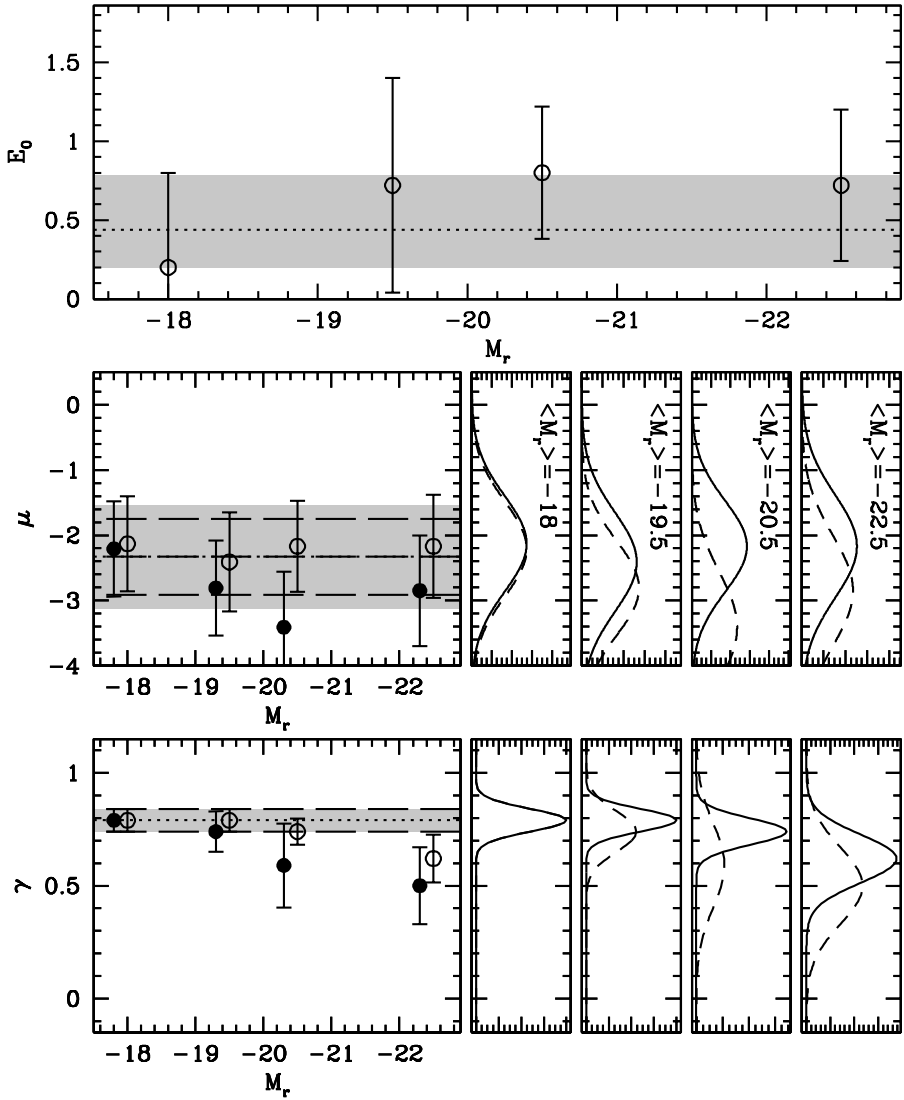,width=16.cm}}
\end{picture}
\caption{
Dependence of the best-fitting parameters on galaxy luminosity, for
spiral galaxies.  
Circles correspond to the best fit parameters in the five-dimensional
parameter space. 
Open symbols show results when including dust, filled
circles, results with no dust.  The horizontal lines indicate the  
best fitting parameters obtained from the full sample of spiral galaxies
(dotted lines show the best fit when including dust, thick dashed with no dust).
Shaded areas and thin dashed lines show the variance in the parameters that best fit the observed
distributions of projected axis ratios.
Top panel: variations in the best-fitting
values of extinction (filled symbols, dashed line).  Error-bars show the ranges
of extinction values such that the likelihood is above $\Delta{\cal L}=0.36$.
Middle panel: variations
in the typical 
$\mu=\log(1-B/A)$
axis ratio; the error bars indicate the best
fitting width $\sigma_{\gamma}$ for the Gaussian
distribution of $\gamma$ values used in the model. The corresponding distributions
for each subsample are shown on the right in individual sub-panels. 
Bottom panel: same as middle panel for 
$\gamma=1-C/B$ (thick dashed lines are not shown to improve clarity).
}
\label{fig:var_expgx}
\end{figure*}

\begin{figure*}
\begin{picture}(430,370)
\put(0,-20){\psfig{file=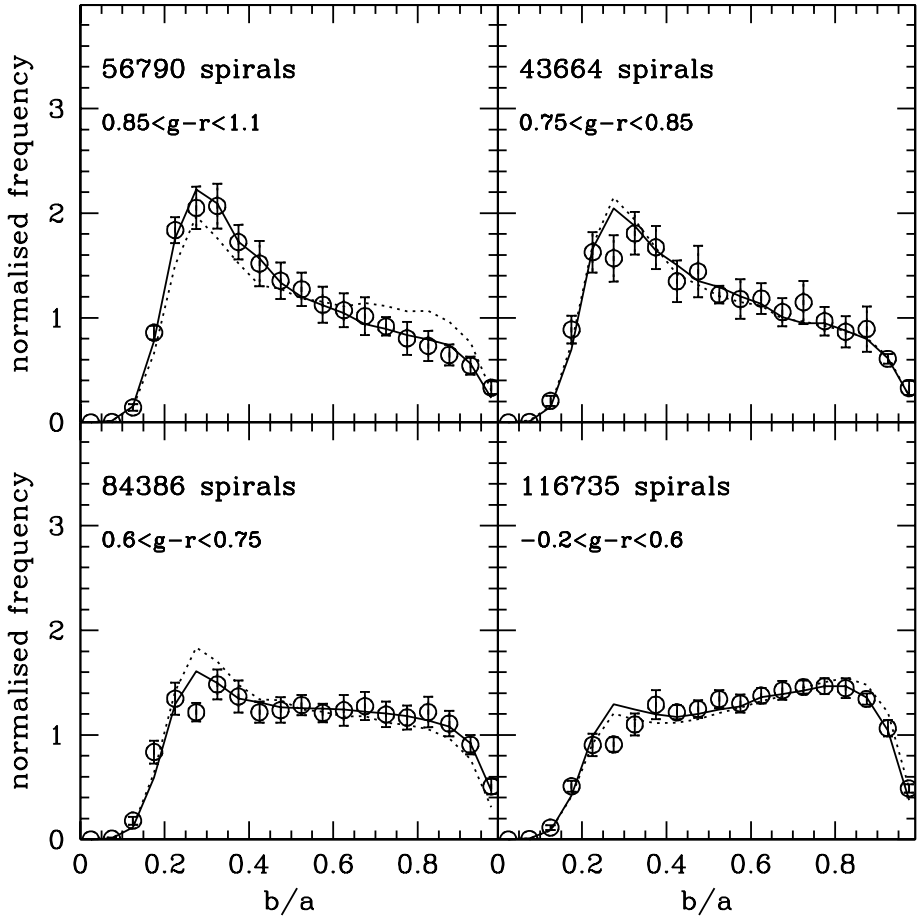,width=14.cm}}
\end{picture}
\caption{
Comparison between the best fit model axis ratio distribution to the full
sample of spiral galaxies for different $g-r$ colour cuts (dotted lines), and the
actual measured distributions from samples of spiral galaxies with different
$g-r$ colours in the SDSS DR6
(open symbols with error bars).  Errors are calculated using the Jack-knife method.
Solid lines show the results of fitting the model to each individual subsample including
dust.  
}
\label{fig:model_dr3c_abscut_expgx}
\end{figure*}

\begin{figure*}
\begin{picture}(430,415)
\put(0,-30){\psfig{file=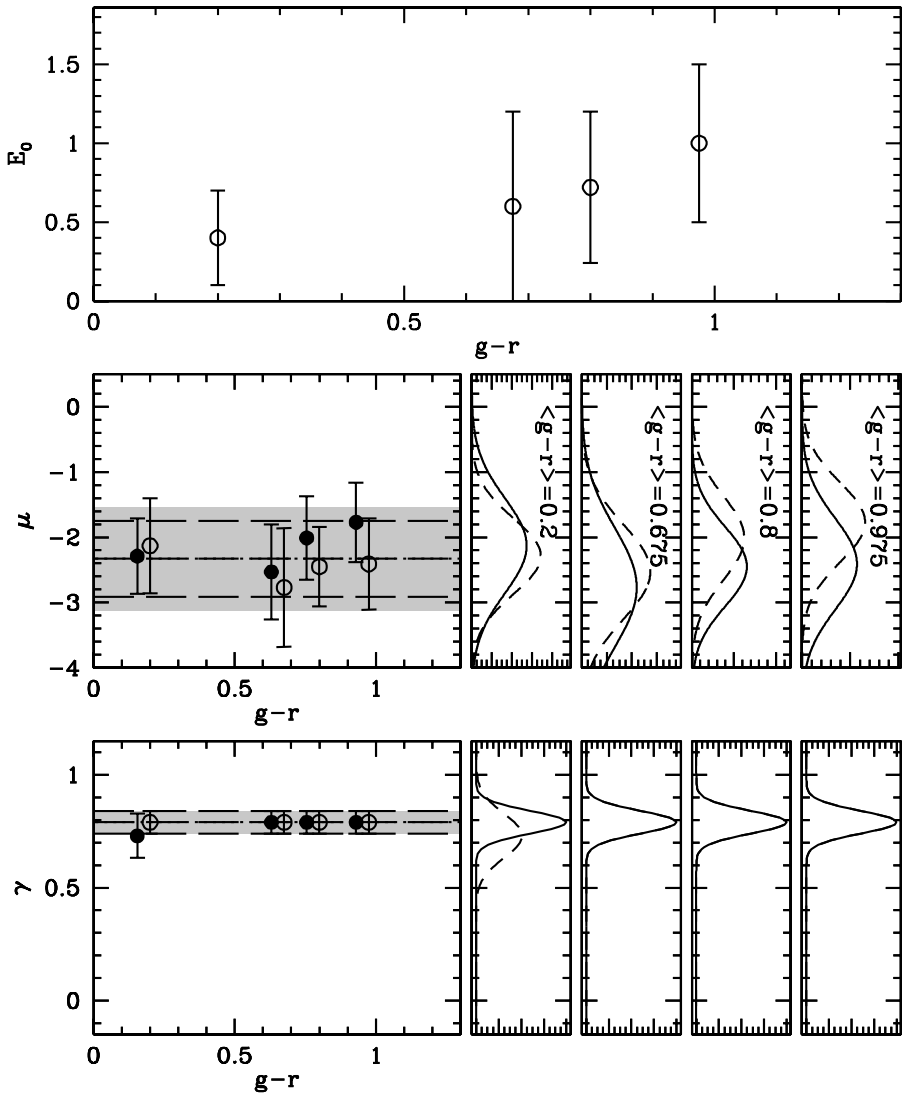,width=16.cm}}
\end{picture}
\caption{
Dependence of the best-fitting parameters on galaxy colour, for
spiral galaxies.  
Panels, lines and symbols are as in Figure \ref{fig:var_expgx}.
}
\label{fig:var_gmr_expgx}
\end{figure*}

\begin{figure*}
\begin{picture}(430,370)
\put(0,-20){\psfig{file=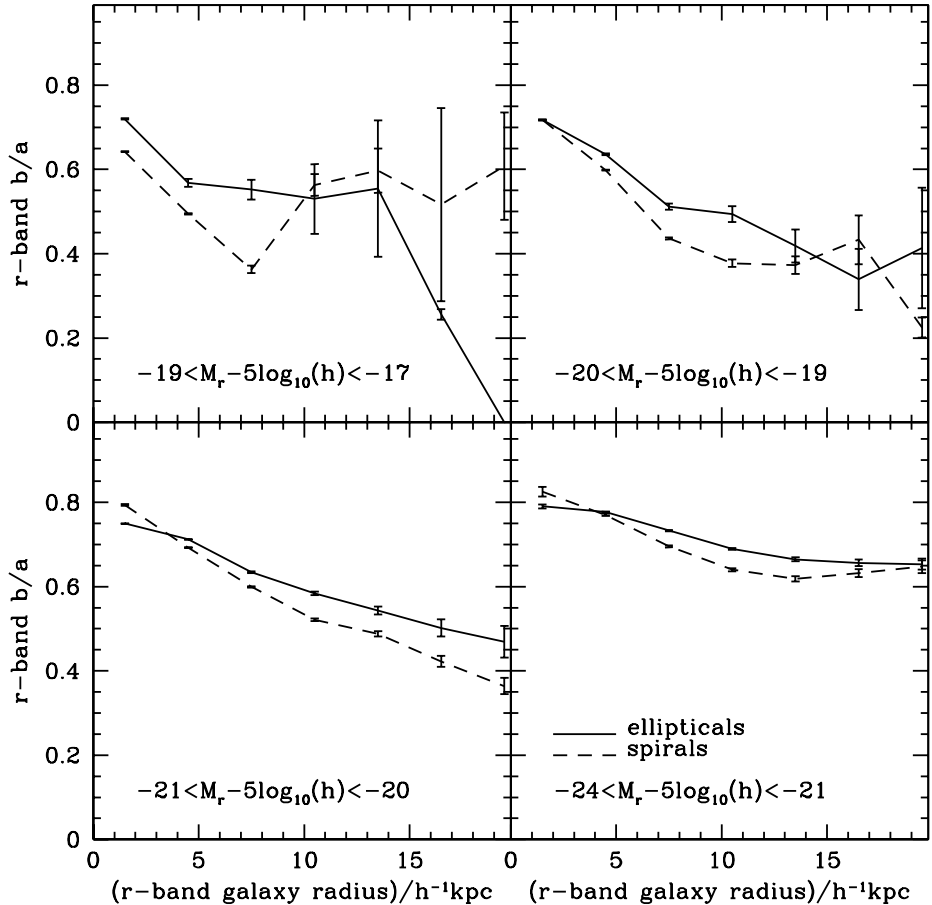,width=14.cm}}
\end{picture}
\caption{
Variation of projected axis ratio $b/a$ with the physical size of galaxies.
Dashed lines correspond to spiral galaxies and
solid lines to elliptical galaxies.  Each panel corresponds to a
different luminosity bin.
}
\label{fig:rcomdr3}
\end{figure*}

\begin{figure*}
\begin{picture}(430,370)
\put(0,-25){\psfig{file=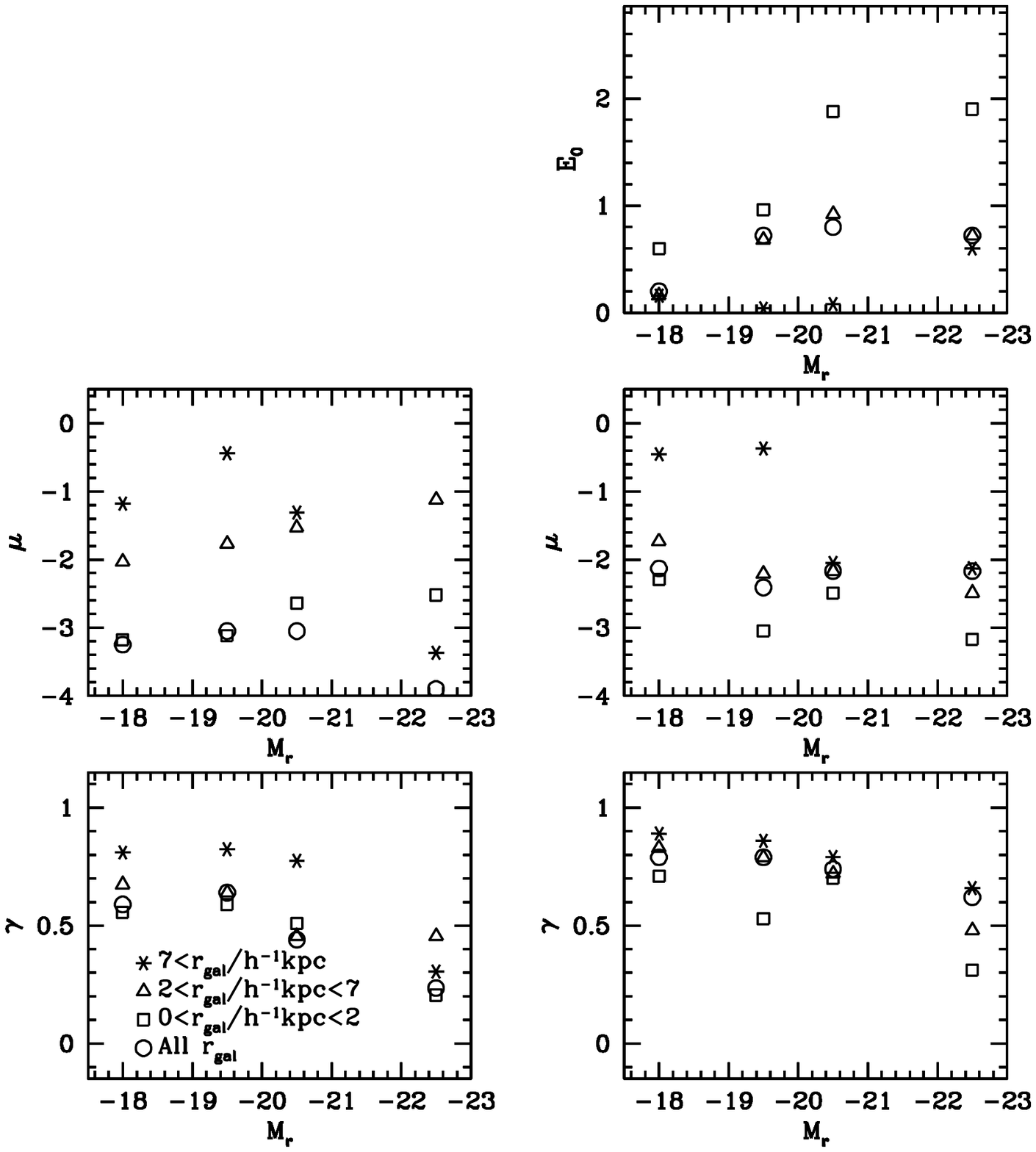,width=14.cm}}
\end{picture}
\caption{
Best-fitting parameters as a function of galaxy luminosity for different galaxy sizes (in
different symbols as shown in the figure key).
Left panels correspond to elliptical galaxies, and right panels to spiral galaxies.
The top panel shows the variation of the extinction, $E_0$, middle panels
show the mean value of $\mu$, and bottom panels show the variation of the $\gamma$ parameter.
}
\label{fig:rcompar}
\end{figure*}

\subsection{Spiral galaxies}

The distribution of spiral galaxy axis ratios depends both on absolute
magnitude and colour (Figure~\ref{fig:absmdr3}), which we interpret as
the effect of dust in the rotational plane of the galaxies.  We model
this as described in \S~\ref{ssec:dust}, giving a five-parameter model.  
We use the grid of parameters from Table \ref{table:grid2} to find the
model quantities that best reproduce the distribution of projected
spiral galaxy shapes.  

The location of the minimum $\chi^2$ is given in
Table~\ref{table:bestfitsexp};  covariance between the parameters in
our model explains the discrepancies with the marginalised values.
As can be seen, in most cases the $\chi^2$ values are small, which indicates an excellent
agreement between the model and the data.

This corresponds to a likelihood more than two orders of magnitude higher than
the best fit found by Ryden (2004) for the spiral galaxies in the Data Release 1
of the SDSS.  Our inclusion of a dust model is partly responsible for
the improvement in the agreement between model and data; the best fit model
with no dust to the full SDSS-DR6 sample of spiral galaxies is characterised
by $\chi^2/{\rm dof}=1.14$, somewhat higher than for the model with
dust, for which  $\chi^2/{\rm dof}=0.41$. 
The sample used by Ryden (2004) is also a factor of $\sim 25$ smaller than our full sample
of spiral galaxies, and is therefore prone to higher sample variance.

It should be noted that dust extinction and the parameter $\mu$ are
somewhat 
degenerate (none of the other parameter pairs show appreciable
degeneracy).  Figure \ref{fig:2Dcontours} shows the marginalised $\mu$
vs. $E_0$ likelihood contours for the full sample of spiral galaxies. 
Interestingly, the axis ratio distribution alone allows the detection
of extinction at only slightly better than $1\,\sigma$.    However, the
strong relationship between colour and axis ratios shown in figure \ref{fig:absmdr3},
and the results by Shao \etal\ (2007), Unterborn \& Ryden (2008) and 
Maller \etal\ (2008), strongly indicate the presence of
dust extinction in spirals.

Figure \ref{fig:model_dr3_expgx} compares the observed spiral axis ratio
distribution with that from our model; the two are in excellent
agreement.  
This figure uses the full
spiral sample; we have assumed that the dust and shape properties of
spirals are independent of luminosity.  We test that assumption in
Figure~\ref{fig:model_dr3_abscut_expgx}, which shows the observed
$b/a$ distribution for spirals in different luminosity ranges, as well
as the model prediction (dashed lines).  The luminosity dependence in this model
comes about solely from the angular selection function from
equation~(\ref{eq:psi}) through the different ranges of luminosity
that define each sample.  
There is excellent agreement between the low luminosity
spirals and the model; indeed, the low-luminosity subsample has a
$b/a$ distribution very close to that of the full sample, which is a
consequence of our $1/V_{max}$ weighting.  This agreement
progressively degrades as we go towards 
higher luminosity.  Indeed, as we saw for ellipticals, the
high-luminosity spirals tend to have larger axis ratios (i.e., to be
more round) than do low-luminosity objects.  

\begin{table}
\centering
\begin{minipage}{80mm}
\caption{Initial parameter grid for Spirals}
\begin{tabular}{@{}cccc@{}}
\hline
Parameter & Min. Value & Max. Value & Number of steps\\
\hline
$E_0$ & $0.0$ & $2.7$ & $14$ \\
$\mu$ & $-4.05$ & $-0.05$ & $21$ \\
$\sigma$ & $0.4$ & $3.0$ & $18$ \\
$\gamma$ & $0.09$ & $0.99$ & $19$ \\
$\sigma_{\gamma}$ & $0.01$ & $0.36$ & $12$ \\
\hline
\label{table:grid2}
\end{tabular}
\end{minipage}
\end{table}

There are two ways we might model this effect.  
A higher dust column for more luminous galaxies could
preferentially remove edge-on objects from the high-luminosity bin, as
suggested by Huizinga \& van Albada (1992).
Alternatively, as in ellipticals, there could be a direct correlation
between three-dimensional shape and luminosity for spirals (e.g.,
Giovanelli \etal\ 1995).  
Indeed, there is a strong correlation between Hubble type and
luminosity whereby high-luminosity spirals tend to be early types with
large bulges (Roberts \& Haynes 1994, Tasca \& White 2005).  This can
explain the low number of small $b/a$ objects at  
such luminosities.

We test these two options by fitting our model separately to galaxies
in each range of absolute magnitude shown in
Figure~\ref{fig:model_dr3_abscut_expgx}; the solid lines show the best
fit distributions with $E_0$ as a free parameter, and the dashed lines the best fits without dust ($E_0=0$).
Figure \ref{fig:var_expgx} shows the 
dependence of the best-fit parameters as a function of absolute magnitude.  
The marginalised estimates are in good agreement and are not shown.
Filled symbols show results with no dust, and open symbols show the results 
when dust extinction is taken into account.  
As in Figure \ref{fig:var_devgx}, error bars indicate the $1-\sigma$ width of the distributions
in $\mu$ and $\gamma$.  The
sub-boxes in the middle and bottom panels
show the model distribution
functions of  $x\equiv B/A$ and $y\equiv C/A$ for each subsample in order of increasing sample luminosity 
from left to right; the model with dust is shown as solid lines, without dust as dashed lines.
As can be seen, the main change
in the inferred parameters is the value of $\mu$, which is systematically lower when dust extinction
is not considered (this is a $1-\sigma$ effect for the two most
luminous subsamples). This can 
also be seen to a lesser extent for the parameter $\gamma$. We find
that the roundness of disks does not change significantly with galaxy
luminosity. 
The more luminous spiral galaxies show higher height to diameter ratios probably due
to the presence of larger galactic bulges.

\begin{table*}
\centering
\begin{minipage}{120mm}
\caption{Best fit model parameters for spiral galaxies: Full sample and dependence on luminosity.}
\begin{tabular}{@{}cccccc@{}}
\hline
Par.  &  Sample 1 &  Sample 2 &  Sample 3 &  Sample 4 & All\\
  & \tiny $-17>M_r>-19$ & \tiny$-19>M_r>-20$ & \tiny$-20>M_r>-21$ & \tiny$-21>M_r>-24$ \\
\hline
$E_0$             & $0.20\pm0.61$   &  $0.72\pm0.68$   & $0.8\pm0.7$   & $0.72\pm0.49$   & $0.44\pm0.24$\\
$\mu$             & $-2.13\pm0.38$  &  $-2.41\pm0.48$ & $-2.17\pm0.41$ & $-2.17\pm0.34$ & $-2.33\pm0.13$\\
$\sigma$          &  $0.73\pm0.19$  &  $0.76\pm0.25$  &  $0.70\pm0.35$  &  $0.79\pm0.31$  & $0.79\pm0.16$\\
$\gamma$          &  $0.79\pm0.02$  &  $0.79\pm0.03$ &  $0.74\pm0.03$ &  $0.62\pm0.04$ & $0.79\pm0.02$\\
$\sigma_{\gamma}$ &  $0.048\pm0.007$&  $0.051\pm0.007$ &  $0.06\pm0.01$ &  $0.11\pm0.02$ & $0.050\pm0.015$\\
$\chi^2/$dof        & $0.21$ & $0.49$ & $1.31$& $0.43 $ & $0.41$ \\
\hline
\label{table:bestfitsexp}
\end{tabular}
\end{minipage}
\end{table*}

\begin{table*}
\centering
\begin{minipage}{100mm}
\caption{Best fit model parameters for spiral galaxies: dependence on galaxy colour.}
\begin{tabular}{@{}ccccc@{}}
\hline
Par.  &  Sample 1 &  Sample 2 &  Sample 3 &  Sample 4 \\
 & \tiny $-0.2<g-r<0.6$ & \tiny$0.6<g-r<0.75$ & \tiny$0.75<g-r<0.85$ & \tiny$0.85<g-r<1.1$ \\
\hline
$E_0$             & $0.4\pm0.3$   & $0.6\pm0.6$   & $0.7\pm0.5$   & $1.0\pm0.5$ \\
$\mu$             & $-2.13\pm0.41$ & $-2.77\pm0.45$ & $-2.45\pm0.38$ & $-2.41\pm0.33$ \\
$\sigma$          &  $0.7\pm0.4$  &  $0.61\pm0.31$  &  $0.91\pm0.36$  &  $0.73\pm0.32$ \\
$\gamma$    &  $0.80\pm0.03$ &  $0.80\pm0.03$ &  $0.80\pm0.02$ &  $0.79\pm0.02$ \\
$\sigma_{\gamma}$ &  $0.054\pm0.012$ &  $0.054\pm0.008$ &  $0.052\pm0.006$ &  $0.050\pm0.005$\\
$\chi^2/$dof        & $0.33$ & $0.17$ & $0.15$& $0.20 $  \\
\hline
\label{table:bestfitsexpc}
\end{tabular}
\end{minipage}
\end{table*}

The inferred intrinsic shapes of spiral galaxies are robust to the effects of dust.
The amount of extinction $E_0$ appears to be independent of
luminosity, but 
the uncertainty in this parameter is large.  The lower values
of $\gamma\simeq 0.6$ characterising bright galaxies results in
high values of projected $b/a$ regardless of the viewing angle.  This, in conjunction
with the effect of extinction to reduce the number of objects seen edge-on,
means that $E_0$ is not very well constrained from this analysis.  

\subsection{Axis Ratio Dependencies on Colour and Size}

We now explore the different intrinsic shapes of galaxies according to their $g-r$ colour.
This analysis may provide better constraints on the dust extinction in spiral galaxies.
Figure \ref{fig:model_dr3c_abscut_expgx} shows the axis ratio distribution in bins of $g-r$.
The dotted lines in this figure show the model using a fixed galaxy shape and dust corresponding
to the best fit to the full spiral galaxy sample.
The agreement is reasonable for all except the reddest subsample.
Therefore, we allow the parameters to vary (solid lines), and the fit to the different colour
subsamples is improved significantly.
Table~\ref{table:bestfitsexpc} shows the best fit parameters for each colour subsample.
Figure \ref{fig:var_gmr_expgx} shows the resulting best fit parameters with and
without dust (open and filled symbols, respectively).  
The intrinsic shapes of galaxies do not show a clear dependence with colour, but
the inferred amount of dust extinction is consistent with an increase 
from $E_0=0.4$ for blue galaxies, to $E_0=1$ for the reddest sample. 
There are only small differences between the recovered values of $\mu$ and $\gamma$ when including
dust.  

Up to this point we had assumed that there is no dependence of shape and dust extinction with
the physical size of the galaxy.  The best-fit parameter values are
those appropriate for galaxies of median size within each subsample. 
However, the intrinsic shape and the amount of dust extinction of
galaxies might depend on their sizes; galaxies of different sizes but
similar luminosities might have
different dynamical histories, and the dust column might reasonably be
larger in larger galaxies.  

The top panel of Figure \ref{fig:rcomdr3} shows the dependence of the median $b/a$ on
projected galaxy size as given by the photometric model scale-length
$r_{gal}$ (exponential or
de Vaucouleurs depending on the value of the $\rm fracDeV$ parameter).  
The median $b/a$ decreases significantly for larger galaxies for
both spirals and ellipticals.  This effect is present at all galaxy
luminosities.  

With this dependence in mind, Figure~\ref{fig:rcompar}
shows the results of fitting our model to samples of galaxies in bins
of physical size at constant luminosity, where the left panels show the results for elliptical galaxies 
(parameters $\mu$ and $\gamma$), and the right panels for spiral galaxies (parameters
$\mu$, $\gamma$, and $E_0$).  
The best-fit
parameters are presented in Table \ref{table:bestfitsradf} for faint, spiral and
elliptical galaxies, and in Table \ref{table:bestfitsradb} for bright galaxies.
At a given luminosity, small elliptical galaxies have similar model
parameters as those of the full sample of ellipticals, although at high
luminosity, they tend to show slightly more elongated shapes (as
reflected in the $\mu$ parameter).  Large elliptical 
galaxies, on the other hand, are more elongated than 
the full sample.  In particular, the $r_{gal}>7\,h^{-1}$ kpc sample can be identified 
with elongated prolate shapes.  
Quantitatively, at low luminosities ($M_r=-18$), the median axis
ratios are $B/A=0.94$ and $C/B=0.42$ for small galaxies, and 
$B/A=0.85$ and $C/B=0.1$ for large galaxies.  We examined the SDSS
images and colours of ellipticals of the most extreme axis ratio
($b/a<0.1$); all appeared to be correctly classified. 

\begin{table*}
\centering
\begin{minipage}{100mm}
\caption{Best fit model parameters: dependence on galaxy size for galaxies with $M_r>-19$. 
Small galaxies satisfy $r_{gal}<2$h$^{-1}$kpc, medium galaxies, $2$h$^{-1}$kpc$<r_{gal}<7$h$^{-1}$kpc, and large galaxies, $r_{gal}>7$h$^{-1}$kpc.}
\begin{tabular}{@{}ccccccc@{}}
\hline
Par.  &  Spirals &  Spirals &  Spirals &  Ellipticals& Ellipticals& Ellipticals \\
\  & Small & Medium  & Large & Small & Medium & Large \\
\hline
$E_0$             & $0.6$   & $ 0.16$ & $ 0.16 $ & & &  \\
$\mu$             & $-2.29$ & $-1.73$ & $-0.45$ & $-3.18$ & $-2.03$ & $-1.18$ \\
$\sigma$          & $0.76$   & $0.64$   & $1.54$   & $0.75$  & $1.6$   & $1.6$   \\
$\gamma$          & $0.71$  & $0.83$  & $0.89$  & $0.445$  & $0.325$ & $0.19$  \\
$\sigma_{\gamma}$ & $0.05$  & $0.02$  & $0.01$  & $0.17$   & $0.17$  & $0.05$  \\
\hline
\label{table:bestfitsradf}
\end{tabular}
\end{minipage}
\end{table*}

\begin{table*}
\centering
\begin{minipage}{100mm}
\caption{Best fit model parameters: dependence on galaxy size for galaxies with $M_r<-21$.  Small, medium and large galaxies are selected as in table \ref{table:bestfitsradf}.}
\begin{tabular}{@{}ccccccc@{}}
\hline
Par.  &  Spirals &  Spirals &  Spirals &  Ellipticals& Ellipticals& Ellipticals \\
  & Small & Medium  & Large & Small & Medium & Large \\
\hline
$E_0$             & $1.90$  & $ 0.72$ & $ 0.59$ & & &  \\
$\mu$             & $-3.17$ & $-2.49$ & $-2.13$ & $-2.52$ & $-1.12$ & $-3.37$ \\
$\sigma$          & $0.91$  & $0.58$  & $0.79$   & $2.7 $  & $2.6$   & $0.85$   \\
$\gamma$          & $0.31$  & $0.48$  & $0.66$  & $0.795$  & $0.545$ & $0.695$  \\
$\sigma_{\gamma}$ & $0.04$  & $0.14$  & $0.09$  & $0.22$   & $0.13$  & $0.17$  \\
\hline
\label{table:bestfitsradb}
\end{tabular}
\end{minipage}
\end{table*}

Low-luminosity spiral galaxies, on the other hand, go from median axis
ratios of $B/A=0.9$ and $C/B=0.3$ for small 
sizes to $B/A=0.36$ and $C/B=0.12$ for large galaxies.  Thus, the disk
thickness decreases with increasing size.  
Shao \etal\ (2007) assumed that $\mu$ and $\sigma$ are independent of
size, whereas we found that most of the changes in the shape
of galaxies with size are actually absorbed by variations in these
parameters.  However,
the conclusions of Shao \etal\ are robust to this detail and indicate, as our results do,
that larger spiral galaxies tend to have flatter disks.

\section{Discussion and Conclusions}
\label{sec:discussion}

In this paper we have addressed the problem of reproducing the observed distributions
of projected axis ratios of galaxies from the SDSS.  We have introduced a number of improvements
over previous works including,  i) the use of larger samples of galaxies made
possible by the introduction of an iterative $1/V_{max}$ weighting scheme,
ii) the inclusion of the effects
of dust extinction on the distribution of apparent $b/a$ axis ratios for spiral galaxies, 
iii) the analysis of
dependence of galaxy shapes on galaxy luminosity, colour, and physical size.

We developed a simple model for the effects of dust extinction on the distribution of 
apparent axis ratios and used it to constrain the intrinsic shapes of spiral galaxies.  
We characterise a given galaxy as a triaxial ellipsoid of axes $A, B$, and $C$ from major 
to minor.  The full sample of spiral galaxies is characterised by the mean and
standard deviation of $1 - C/B$ of $\mu=-2.33\pm0.13$ and
$\sigma=0.79\pm0.16$; the distribution of $\log(1-B/A)$ is modeled as
a lognormal with mean $\gamma=0.79\pm0.02$ and standard deviation
$\sigma_{\gamma}=0.050\pm0.015$.  These values are in good agreement
with the thickness of disks derived by Ryden (2004), who finds
$\gamma=0.216$, although her face-on disks are more elliptic, $\mu=-1.85$, and the distribution
widths are somewhat different from ours.
More recently, Unterborn \& Ryden (2008) analyse the shapes of
a sample of spiral galaxies which has been corrected for the biases
introduced by dust on the distribution of projected shapes.  The resulting
spiral shapes are characterised by $\mu=-2.56$, $\sigma=0.91$, $\gamma=0.216$ and
$\sigma_{\gamma}=0.067$, which are roughly consistent with the results
we present here; however, their sample selection is explicitly dependent
on their model for the dependence of dust on inclination angle, which makes it difficult 
to make a more quantitative comparison. 

We also studied variations in the intrinsic shapes of galaxies with galaxy luminosity 
(Table \ref{table:bestfitsexp}), colour(Table \ref{table:bestfitsexpc}), and physical
size (Tables \ref{table:bestfitsradf} and \ref{table:bestfitsradb}).  
As luminosity and colour are correlated, and in particular more luminous spiral 
galaxies show larger bulges, we find rounder spiral galaxies at larger luminosities and $g-r$
colours.  At a given luminosity, larger galaxies tend to be flatter. 

Our quoted parameters are obtained by marginalising over the extinction
parameter, $E_0$.  The extinction is not well-constrained by the
axis-ratio distribution alone, but for our full sample, we find a
value $E_0=0.44\pm0.24$ after marginalising over other parameters, in
good agreement with results from the literature.  

\begin{figure}
\begin{picture}(230,220)
\put(0,0){\psfig{file=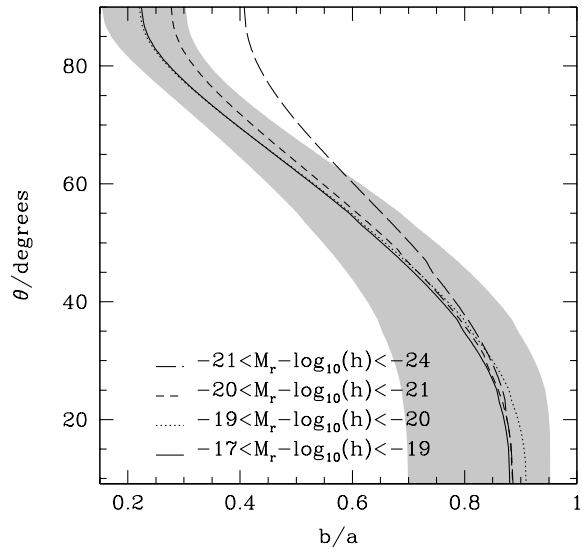,width=8.cm}}
\end{picture}
\caption{
Relation between the polar viewing angle, $\theta$, and projected axis
ratios, $b/a$, for subsamples of spiral galaxies of different
absolute magnitudes.  The shaded area encloses $80\%$ of the possible
viewing angles associated with a given value of $b/a$.
}
\label{fig:badetheta}
\end{figure}

Using our results from Table \ref{table:bestfitsexp}, we determine the
relation between the observed projected axis ratios of spiral galaxies
and the inclination angles of their disks for four different
luminosity ranges.  In order to do this, we simply calculate
$(b/a)_{max}$ at which the distribution of projected axis ratios
peaks, as well as the $10\%$ and $90\%$ percentiles, for a given narrow range
of viewing angles.  We use the percentiles to infer the range of
angles that correspond to a given $b/a$ ratio that can be measured
from a spiral galaxy.  Figure \ref{fig:badetheta} and Table
\ref{table:badetheta} show the relation between projected axis ratio
$b/a$ and polar viewing angle $\theta$ for
spiral galaxies of different luminosities.  The shaded area shows the
ranges of polar viewing angles that correspond to a given value of
$b/a$ for the fainter sample.  As brighter galaxies tend to have
thicker disks, the value of $b/a$ corresponding to edge-on bright
galaxies is higher than for fainter galaxies.  Note that these results
take into account the effects of dust extinction.

\begin{table*}
\centering
\begin{minipage}{100mm}
\caption{Relation between $b/a$ and polar viewing angle, $\theta$, for subsamples
of spiral galaxies corresponding to different ranges of absolute magnitude}
\begin{tabular}{@{}ccccc@{}}
\hline
$\frac{\theta}{\rm degrees}$ & $b/a$ & $b/a$ & $b/a$ & $b/a$\\
 & $\langle M_r\rangle=-18.0$ & $\langle M_r\rangle=-20.1$ & $\langle M_r\rangle=-21.2$ & $\langle M_r\rangle=-22.7$\\
\hline
$           3$&$      0.8815$&$      0.9104$&$       0.886$&$      0.8864$\\
$           7$&$      0.8812$&$        0.91$&$      0.8858$&$      0.8863$\\
$          11$&$        0.88$&$      0.9083$&$      0.8847$&$      0.8855$\\
$          15$&$      0.8783$&$      0.9059$&$      0.8833$&$      0.8831$\\
$          19$&$      0.8724$&$       0.898$&$      0.8778$&$      0.8777$\\
$          23$&$      0.8616$&$      0.8844$&$      0.8673$&$      0.8731$\\
$          27$&$      0.8487$&$      0.8687$&$      0.8513$&$      0.8603$\\
$          31$&$      0.8265$&$      0.8429$&$      0.8343$&$       0.842$\\
$          35$&$      0.7981$&$      0.8115$&$       0.807$&$      0.8188$\\
$          39$&$      0.7687$&$      0.7797$&$      0.7745$&$      0.7911$\\
$          43$&$        0.73$&$      0.7388$&$      0.7373$&$      0.7596$\\
$          47$&$      0.6871$&$       0.694$&$      0.6963$&$      0.7308$\\
$          51$&$      0.6407$&$      0.6458$&$      0.6566$&$      0.6929$\\
$          55$&$      0.5946$&$      0.5983$&$      0.6091$&$      0.6531$\\
$          59$&$      0.5424$&$      0.5449$&$      0.5597$&$      0.6124$\\
$          63$&$      0.4885$&$      0.4899$&$      0.5092$&$      0.5721$\\
$          67$&$      0.4339$&$      0.4344$&$      0.4587$&$      0.5332$\\
$          71$&$      0.3801$&$      0.3797$&$      0.4099$&$      0.4968$\\
$          75$&$      0.3291$&$      0.3279$&$      0.3647$&$      0.4647$\\
$          79$&$      0.2838$&$      0.2816$&$      0.3258$&$      0.4383$\\
$          83$&$      0.2484$&$      0.2454$&$      0.2964$&$      0.4195$\\
$          87$&$      0.2284$&$      0.2248$&$      0.2805$&$      0.4095$\\
\hline
\label{table:badetheta}
\end{tabular}
\end{minipage}
\end{table*}

The axis ratio distribution of elliptical galaxies shows no dependence
on colour, suggesting that dust extinction is not important for this
sample, and we do not include it in our modelling.  The full sample of elliptical
galaxies are characterised by parameters $\mu=-2.2\pm0.2$, $\sigma=1.4\pm0.10$, $\gamma=0.57\pm0.06$ and
$\sigma_{\gamma}=0.21\pm0.02$, which correspond to slightly oblate spheroids in agreement with
results by Vincent \& Ryden (2005).  More luminous ellipticals tend to
be rounder, although ellipticals are oblate at all luminosities.
Although Vincent \& Ryden (2005) found that the lowest luminosity
ellipticals are best fitted by prolate spheroids, their inferred $C/A$
axis ratio is consistent with our estimates.   

This work presented the most detailed statistical study of shapes of
galaxies separated in spirals and ellipticals to date, and includes
the effects of dust on the distribution of projected shapes of
spirals in a self-consistent way. 
It is remarkable that dust can be
inferred from the distribution of projected axis ratios alone,
although only at a low statistical significance.  
The dependence of intrinsic galaxy shapes and extinction 
with luminosity, colour and size expands the range of tests that galaxy
formation models need to satisfy, to continue improving the modelling
of the processes that drive the evolution of galaxies.

\section*{Acknowledgments}
This work was supported in part by the FONDAP ``Centro de Astrof\'\i sica" and Fundaci\'on
Andes.  NP was supported by a Proyecto Fondecyt Postdoctoral
no. 3040038 and Regular No. 1071006, and MAS was supported by NSF grants
AST-0307409 and AST-0707266.  We have benefited from helpful
discussions with Diego Garc\'\i a Lambas 
and Jeremiah P. Ostriker.  

    Funding for the SDSS and SDSS-II has been provided by the Alfred
    P. Sloan Foundation, the Participating Institutions, the National
    Science Foundation, the U.S. Department of Energy, the National
    Aeronautics and Space Administration, the Japanese Monbukagakusho,
    the Max Planck Society, and the Higher Education Funding Council
    for England. The SDSS Web Site is http://www.sdss.org/. 

    The SDSS is managed by the Astrophysical Research Consortium for
    the Participating Institutions. The Participating Institutions are
    the American Museum of Natural History, Astrophysical Institute
    Potsdam, University of Basel, University of Cambridge, Case
    Western Reserve University, University of Chicago, Drexel
    University, Fermilab, the Institute for Advanced Study, the Japan
    Participation Group, Johns Hopkins University, the Joint Institute
    for Nuclear Astrophysics, the Kavli Institute for Particle
    Astrophysics and Cosmology, the Korean Scientist Group, the
    Chinese Academy of Sciences (LAMOST), Los Alamos National
    Laboratory, the Max-Planck-Institute for Astronomy (MPIA), the
    Max-Planck-Institute for Astrophysics (MPA), New Mexico State
    University, Ohio State University, University of Pittsburgh,
    University of Portsmouth, Princeton University, the United States
    Naval Observatory, and the University of Washington.

\bsp

\label{lastpage}

\end{document}